\documentclass[11pt]{article}
\usepackage{amsfonts}
\usepackage{amssymb}
\usepackage{mathrsfs}
\usepackage{graphicx}
\usepackage{amsmath,amsthm,amssymb,amscd}
\usepackage[all]{xy}
\usepackage{subfig}
\usepackage{hyperref}
\usepackage{multirow}

\usepackage[dvips]{color}
\usepackage{amsfonts}
\usepackage{float}
\newcommand{\ca}{{\cal A}}

\newcommand{\cd}{{\cal D}}

\newcommand{\ck}{{\cal K}}
\newcommand{\cn}{{\cal N}}
\newcommand{\cm}{{\cal M}}
\newcommand{\cl}{{\cal L}}

\newcommand{\cj}{{\cal J}}
\newcommand{\cg}{{\cal G}}
\newcommand{\ch}{{\cal H}}

\newcommand{\crr}{{\cal R}}
\newcommand{\co}{{\cal O}}
\newcommand{\cp}{{\cal P}}
\newcommand{\cs}{{\cal S}}
\newcommand{\cq}{{\cal Q}}

\newcommand{\cw}{{\cal W}}
\newcommand{\cx}{{\cal X}}

\newcommand{\nn}{\nonumber}
\def\eqa{\begin{eqnarray}}
\def\eqae{\end{eqnarray}}
\def\eq{\begin{equation}}
\def\eqe{\end{equation}}
\def\be{\begin{equation}}
\def\ee{\end{equation}}
\def\bea{\begin{eqnarray}}
\def\eea{\end{eqnarray}}
\def\ba{\begin{array}}
\def\ea{\end{array}}
\def\bd{\begin{displaymath}}
\def\ed{\end{displaymath}}

\def\>{\rangle}
\def\<{\langle}
\def\a{\alpha}
\def\b{\beta}

\def\del{\delta}
\def\e{\epsilon}
\def\f{\phi}
  
\def\g{\gamma}
\def\h{\eta}

\def\k{\kappa}
\def\l{\lambda}
\def\m{\mu}
\def\n{\nu}
\def\o{\omega}  
\def\p{\pi}
\def\q{\theta}
\def\r{\rho}
\def\s{\sigma}

\def\G{\Gamma}

\def\L{\Lambda}

\def\S{\Sigma}

\def\pa{\partial}

\newcommand{\bc}{{\mathbb{C}}}
\newcommand{\br}{{\mathbb{R}}}

\def\hl{\hat L}
\def\hm{\hat M}
\def\hn{\hat N}
\def\hj{\hat J}
\def\hx{\hat X}
\def\hg{\hat G}

\def\hhl{\hat{\hat L}}
\def\hhj{\hat{\hat J}}
\def\hhg{\hat{\hat G}}

\def\hhlm{\hat{\hat \l}}

\newcommand{\fn}{\footnotemark\footnotetext}

\newcommand{\smbox}[1]{\ \mbox{#1}\ }

\numberwithin{equation}{section}

\begin{document}

\begin{titlepage}

\begin{flushright}
MCTP 12-31 \\
\end{flushright}
\vspace{1cm}

\begin{center}
{\Large \bf Dualities from higher-spin supergravity}\\[1cm]
Cheng Peng~\\[6mm]

{\it Michigan Center for Theoretical Physics}\\
{\it Department of Physics, University of Michigan}\\
{\it Ann Arbor, MI 48109, USA}\\
~\\
\today
~\\

\end{center}

\vspace{1cm} \centerline{\bf Abstract} \vspace*{0.5cm}

We study the vacuum structure of spin-3 higher-spin supergravity in AdS$_{3}$ spacetime. The theory can be written as a Chern-Simons theory based on the Lie superalgebra $sl(3|2)$. We find three distinct AdS$_{3}$ vacua, AdS$^{(1)}$, AdS$^{(2)}$ and AdS$^{(\rm p)}$, each corresponding to one embedding of the $osp(1|2)$ subalgebra into the $sl(3|2)$ algebra. We explicitly construct the RG flows from AdS$^{(1)}$ to AdS$^{(\rm p)}$ and from AdS$^{(2)}$ to AdS$^{(\rm p)}$,  which identifies AdS$^{(\rm p)}$ as an IR vacuum and AdS$^{(1)}$, AdS$^{(2)}$ as two different UV vacua. Thus a duality is found between the two UV theories in the sense that the two theories, each with a chemical potential turned on, flow to the same IR theory.
Moreover,  we identify a similar structure in the Hamiltonian reductions of the 2d Wess-Zumino-Witten (WZW) model with $sl(3|2)$-valued currents by matching the chiral symmetries there with the asymptotic symmetries of the three different embeddings. Our computation gives an RG interpretation of (certain types of) the Hamiltonian reductions. In addition, it gives a hint of a duality between the 3d higher-spin supergravity and some conformally extended super-Toda theory as suggested by Mansfield and Spence for the bosonic case.

\end{titlepage}

\tableofcontents

%%%%%%%%%%%%%%%%%%%%%%
\section{Introduction}
%%%%%%%%%%%%%%%%%%%%%%

The AdS/CFT duality with higher-spin fields on both sides has drawn a lot of attention in recent years. A long-term effort of Fronsdal, Fradkin, Vasiliev and collaborators has succeeded in constructing gravitational theories with spin $s>2$ gauge fields in arbitrary dimensions \cite{Fronsdal, vasiliev}. These higher-spin theories fit in the context of holography very well. In particular, there are two types, namely A-type and B-type, of parity invariant higher-spin theories in  AdS$_{4}$ spacetime.   Klebanov and Polyakov conjectured a duality between the A-type minimal bosonic higher-spin theory in AdS$_4$ spacetime and the 3d $O(N)$ vector model  \cite{klp}. Sezgin and Sundell proposed a duality between the B-type minimal bosonic higher-spin theory in AdS$_4$ spacetime and the 3d Gross-Neveu model \cite{ss}. In the same paper \cite{ss}, the $\cn=1$ supersymmetric versions of these dualities were also conjectured.  One exciting feature of higher-spin theory is its potential link with the tensionless limit of string theory, as speculated for a long time since \cite{sundborg,ss2}. Recently, the authors of  \cite{gmptwy,cmsy} proposed a generalized duality between parity violating higher-spin theory and the (supersymmetric) Chern-Simons matter theory based on the observation in \cite{agy}, which builds a direct connection to the type IIA string theory. Significant progress has been made on higher-spin holography. For example, exact or slightly broken higher-spin symmetry were shown to impose strong constraints on the CFT by Maldacena and Zhiboedov \cite{mz1,mz2}. Aharony, Gur-Ari and Yacoby clarified the interpolation between the A-type higher-spin theory and the B-type theory \cite{agy2, aggmy}. Different attempts to understand physical origin of the higher-spin holography have been carried out in \cite{dj,jjy,dmr, v12}.  A de-Sitter/CFT higher-spin holography is also conjectured in \cite{ahs,ns,ddjy}. For a recent review, see \cite{gy} and the references therein.

At one-dimension lower, higher-spin theories have also been studied intensively in the context of AdS$_{3}$/CFT$_{2}$ duality.
Gaberdiel and Gupakumar proposed a duality between bosonic higher-spin gravity in AdS$_{3}$ and 2-dimensional $W_{N}$ minimal models \cite{gg}. This duality is refined later in \cite{gg12,ppr} and extended to even spin higher-spin theory \cite{cgkv}. The supersymmetric generalizations have been carried out by   Creutzig, Hikida and Ronne \cite{chr,chr1} and refined by Candu and Gaberdiel \cite{cg,cg2}. This duality has been studied and checked intensively, including the match of the partition functions \cite{gghr,chr,chr1},  the asymptotic symmetries \cite{hr, cfpt, gh,hlpr,hp,ahn1,ahn2}, and the correlation functions \cite{chr2,mz}.  Classical solutions such as conical singularities \cite{chjm,cggr} and black holes with higher spin charges \cite{gk,kp, ghj, tan,dd} are also constructed. Besides, non-AdS holography with higher-spin gauge fields was studied in \cite{ggr}. For recent reviews, see \cite{gg2,agkp1} and the references therein.

An interesting phenomenon observed by Ammon, Gutperle, Kraus and Perlmutter in \cite{agkp} is that $sl(3,\br)$ higher-spin gravity admits two distinct AdS$_3$ vacua with different asymptotic symmetries. The two vacua are obtained from different $sl(2,\br)$ embeddings into $sl(3,\br)$. A holographic RG flow triggered by a finite chemical potential connects the vacuum corresponding to the diagonal embedding in the UV and the principal embedding vacuum in the IR. A detailed relation between the operators in the UV theory and the operators in the IR theory is obtained by perturbing the RG flow and solving the linearized equation of motion. An extension of this work to the thermodynamic properties of higher-spin black holes is carried out in \cite{dfk}.

This paper extends the analysis to the supersymmetric setting. We consider the 3d spin-3 supergravity that can be described as a Chern-Simons theory with $sl(3|2)$ gauge fields.
The supergravity sector is represented by an $osp(1|2)$ subalgebra.  It turns out that there are three different ways to embed an $osp(1|2)$ into the defining $sl(3|2)$. The vacuum solutions corresponding to these 3 embeddings are labeled as AdS$^{(1)}$, AdS$^{(2)}$ and AdS$^{(\rm p)}$.\fn{In this paper, we will abuse the notation AdS$^{(\cdot)}$ for both the higher-spin theory with a certain $osp(1|2)$ embedding and its AdS vacuum solution.} We analyze the vacuum structure of each embedding and explicitly construct a holographic RG flow from AdS$^{(1)}$ to AdS$^{(\rm p)}$ and from AdS$^{(2)}$ to AdS$^{(\rm p)}$, respectively. In this sense, we identify  AdS$^{(\rm p)}$  as an IR theory and the AdS$^{(1)}$, AdS$^{(2)}$ as two different UV theories. Thus a duality is found between the two UV theories in the sense that the two theories, each with a chemical potential turned on, flow to the same IR theory. Moreover, this structure is very similar to the known structure of the Hamiltonian reduction of the 2d Wess-Zumino-Witten (WZW) model with $sl(3|2)$-valued currents. It has been shown by Ahn, Ivanov and Sorin \cite{ais} that three different reductions of the $sl(3|2)$ WZW model exist, each containing the usual Virasoro algebra as a subalgebra of its chiral algebra. They have also shown that one of the three reductions can be obtained from the other two by secondary Hamiltonian reductions. We find an exact match between the chiral symmetries of these three resulting theories and the asymptotic symmetries of the three different embeddings. This gives a hint of a duality between the 3d higher-spin supergravity and some extended version of the Hamiltonian reduced 2d WZW models, as argued in \cite{mansfield,ms} for the bosonic case. In addition, our analysis suggests a physical interpretation of the Hamiltonian reduction procedures as RG flows.
%%%%%%%%%%%%%%%%%%%%%%%%%
%%%%%%%%%%%%%%%%%%%%%%
\section{3d higher-spin gravity, its AdS vacua and the RG flow}\label{review}
%%%%%%%%%%%%%%%%%%%%%%
%%%%%%%%%%%%%%%%%%%%%%

In this section, we review the known results about the higher-spin gravity in 3-dimensional AdS spacetime. We then briefly summarize the result in \cite{agkp}, namely the two distinct AdS vacua that correspond to the two different $sl(2,\br)$ embeddings into $sl(3,\br)$ and the  RG flow between them.

%%%%%%%%%%%%%%%%%%%%%%%%%
\subsection{Higher-spin gravity as a Chern-Simons theory}
%%%%%%%%%%%%%%%%%%%%%%%%%

It is shown in \cite{w88, Witten:2007kt} that 3-dimensional gravity theory with a negative cosmological constant can be reformulated as two copies of Chern-Simons gauge theories  \begin{equation} \label{action1}
I_{EH} = I_{CS}(A,k_{CS}) - I_{CS}(\tilde{A},k_{CS})\;, \qquad
I_{CS} (A,k_{CS})= \frac{k_{CS}}{4 \pi}  \int_{\cm} \mbox{Tr}(A \wedge d A + \frac{2}{3} A \wedge A \wedge A)\;,
\end{equation}
where $A$ is the gauge connection that evaluates in $sl(2,\br)$. The ``Tr'' stands for the trace.

We denote the generators of the Lie algebra $sl(2,\br)$ as $L_{0}, L_{\pm 1}$. Then the  connection expands like
\begin{equation}\label{sl2gm}
A=\sum\limits_{m=-1}^{1}A^{m}L_{m}=\sum\limits_{m=-1}^{1}\sum\limits_{\m=0,\pm}A_{\m}^{m}L_{m}\, dx^{\m}
\,, \quad \tilde A=\sum\limits_{m=-1}^{1} \tilde A^{m}L_{m}=\sum\limits_{m=-1}^{1}\sum\limits_{\m=0,\pm} \tilde A^{m}_{\m}L_{m}\, dx^{\m}
\,.
\end{equation}
The component fields $A_{\m}^{m}, \tilde A_{\m}^{m}$ are related to the vielbeins $e_{\m}^{m}$ and spin connections $\o_{\m}^{m}$ as
\begin{equation}\label{gaugefields}
A_{\m}^{m}= \omega_{\m}^{m} + \frac{1}{\ell} e_{\m}^{m}\;, \qquad\qquad
\tilde{A}_{\m}^{m} = \omega_{\m}^{m} - \frac{1}{\ell} e_{\m}^{m}\,,
\end{equation}
where $m=0,\pm 1$ label generators of the $sl(2,\br)$ algebra, $\m=0,\pm $ are spacetime indices. $\ell$ is a constant with mass dimension -1. After plugging \eqref{gaugefields}, \eqref{sl2gm} into  \eqref{action1} and  imposing the equation of motion, we recover the 3d Einstein-Hilbert  action for pure gravity with the following identification \cite{w88, Witten:2007kt}
\begin{equation}\label{kl}
k_{CS}=\frac{\ell}{8G_3}\frac{1}{(-{\mbox{Tr}} L_{0}^{2})}\,,\qquad \L=-\frac{2}{\ell^{2}}\,.
\end{equation}
where $\L$ is the cosmological constant. For AdS space, we have $\L<0$. With this identification, we regard $|\ell|=\sqrt{-{2}/{\L}}$ as the characteristic AdS radius.
Note that in the convention of \cite{cfpt}, $\mbox{Tr}(L_a L_b)=\frac{1}{2} \h_{ab}$, so $\mbox{Tr}(L_0^2)=-\frac{1}{2}$. Our relation \eqref{kl} thus gives $k_{CS}=\frac{\ell}{4 G_3}$ which agrees with the result in \cite{cfpt}. Meanwhile, the spacetime metric $g_{\m\n}$ is related to the vielbeins by
\begin{equation}\label{genmetric}
g_{\m\n}=\frac{1}{\mbox{Tr} L_{0}^{2}}\mbox{Tr} (e_{\m} e_{\n})=\frac{\ell^{2}}{4 \,\mbox{Tr} L_{0}^{2}}\mbox{Tr} ((\tilde A_{\m}- A_{\m})(\tilde A_{\n}- A_{\n}))\,,
\end{equation}
where $e_{\m}=\sum\limits_{m=0,\pm 1}e_{\m}^{m} L_{m}$.

Generalization to higher-spin theory is straightforward in the Chern-Simons language: we simply replace the $sl(2,\br)$ algebra by a higher rank algebra $\ch$. The simplest case is $\ch=sl(3,\br)$, which is  the spin-3 gravity considered in \cite{agkp}.  The action of this $sl(3,\br)$ higher-spin theory is similar to \eqref{action1}
\begin{equation} \label{action2}
I_{HS} = I_{CS}(\G,k_{CS}) - I_{CS}(\tilde{\G},k_{CS})\;, \qquad
I_{CS} (\G,k_{CS})= \frac{k_{CS}}{4 \pi}  \int_{\cm} \mbox{Tr}(\G \wedge d \G + \frac{2}{3} \G \wedge \G \wedge \G)\;,
\end{equation}
where the gauge connections are evaluated in the $sl(3,\br)$ algebra
\begin{equation}\label{sl3gm}
\G=\sum\limits_{i=-1}^{1}A^{(2)}_{i}L^{(2)}_{i}+\sum\limits_{i=-2}^{2}A^{(3)}_{i}W^{(3)}_{i}
\,, \qquad\tilde \G=\sum\limits_{i=-1}^{1}\tilde{A}^{(2)}_{i}L^{(2)}_{i}+\sum\limits_{i=-2}^{2}\tilde{A}^{(3)}_{i}W^{(3)}_{i}
\,,
\end{equation}
where $L_i^{(2)},\, W_i^{(3)}$ generate the spin-$2$ and spin-$3$ representations respectively and $A_i^{(2)},\, A_i^{(3)},\,\tilde A_i^{(2)},\, \tilde A_i^{(3)}$ are corresponding components of the gauge connections.

The metric in higher-spin theory is analogous to \eqref{genmetric}
\begin{equation}\label{hgenmetric}
g_{\m\n}=\frac{\ell^{2}}{4 \,\mbox{Tr} (L^{(2)}_{0})^{2}}\mbox{Tr} ((\tilde \G_{\m}- \G_{\m})(\tilde \G_{\n}- \G_{\n}))\,.
\end{equation}

%%%%%%%%%%%%%%%%%%%%%%%%%
\subsection{$sl(2,\br)$ embeddings, AdS vacua and  RG flow }\label{rvbos}
%%%%%%%%%%%%%%%%%%%%%%%%%

In this subsection, we review the result in \cite{agkp}. In the Chern-Simons language, the gauge connections that correspond to the AdS vacuum read:
\begin{eqnarray}
\G_{\text{AdS}}=e^{\r} L_{1}dx^{+}+L_{0}d\r\,,  \qquad\tilde{\G}_{\text{AdS}}=-e^{\r} L_{-1}dx^{-}-L_{0}d\r\,, \label{adsvcmb}
\end{eqnarray}
where $L_{0,\pm1}$ are the $sl(2,\br)$ generators and we omit the   superscript $(2)$ in $L^{(2)}_{i}$. Then from \eqref{genmetric} we can compute the metric of the corresponding geometric
\begin{equation}
ds^{2}=d\r^{2}-e^{2\r} dx^{+}dx^{-}\,,
\end{equation}
which represents an AdS$_{3}$  spacetime.

This simple computation tells us that the properties, most importantly the trace structure, of the $sl(2,\br)$ generators $L_{0,\pm1}$ determines the metric of the geometry.  Since it is known that there can be different ways to embed an $sl(2,\br)$ algebra to a higher rank Lie algebra, it is natural to ask what happens to the geometry if we consider different $sl(2,\br)$ embeddings.

For the special case $sl(3,\br)$, there are two different $sl(2,\br)$ embeddings: a principal embedding and a non-principal which is usually called the ``diagonal" embedding. The work \cite{agkp} showed that the gauge connections \eqref{adsvcmb} corresponding to the principal embedding give an AdS$_{3}$ vacuum with unit radius, which we denote as AdS$^{(\rm p)}$ with p standing for ``principal". The gauge connections corresponding to the diagonal embedding were shown to describe an AdS$_{3}$ vacuum with radius $1/2$. We denote it as AdS$^{(\rm d)}$ where d stands for ``diagonal".

We can further consider perturbation solutions around any AdS$_{3}$ vacuum that goes back to the vacuum solution asymptotically. For instance, we can consider the perturbed asymptotic AdS$^{(\rm p)}$ solution
\begin{equation}\label{pertvac}
\G=\bigg(e^{\r} L_{1}-\frac{2\p}{k_{CS}}e^{-\r}\cl(x^{+})L_{-1}-\frac{\p}{2 k_{CS}}e^{-2\r}\cw(x^{+})W_{-2}\bigg)dx^{+}+L_{0}d\r\,,
\end{equation}
where $\cl(x^{+}),\cw(x^{+})$ are asymptotic fields. Due to  the topological nature of the Chern-Simons action, much of the information about this perturbation is encoded in the asymptotic symmetry which is the gauge symmetry that  preserves the form of \eqref{pertvac}. Following the procedure shown in \cite{cfpt,gh}, the asymptotic symmetry of the principal embedding AdS$^{(\rm p)}$ can be shown to be the $W_{3}$ algebra, which contains a spin-2 generator and a spin-3 generator. The authors of \cite{agkp} showed that the asymptotic symmetry of the AdS$^{(\rm d)}$ vacuum is the $W_{3}^{(2)}$ algebra, which contains a spin-2 generator, 2 spin-$3/2$ generators and one spin-1 generator.

Another interesting result in \cite{agkp} is the RG flow from AdS$^{(\rm d)}$ to AdS$^{(\rm p)}$. This flow was explicitly constructed as an interpolation solution between the two vacua \begin{equation}\label{bsint}
A=\l\,e^{\r} L_{1} dx^{+}+e^{2\r}\hl_{1} dx^{-}+L_{0}d\r\,,
\end{equation}
where $L_{1}~ (\hl_{1})$ are the spin-2 generators of the principal (diagonal) embedding. A similar expression exists for the $\tilde{A}$ connection. This solution approaches to the AdS$^{(\rm d)}$ in the $\r\to\infty$ limit, while it approaches to the AdS$^{(\rm p)}$ in the $\r\to-\infty$ limit. In this sense, we get an RG flow from the AdS$^{(\rm d)}$ vacuum in the UV to the AdS$^{(\rm p)}$ vacuum in the IR.

The authors of \cite{agkp} also found the relations between UV fields and IR fields. This is done by adding both the UV and IR perturbations
\begin{eqnarray}
\delta \G &=& \left(\cl_{IR}e^{-\rho} L_{-1} + \cw_{IR} e^{-2\rho} W_{-2} \right)dx^+  \\
&& + \left( J_{UV} W_0 + G^{(1)}_{UV}e^{-\rho}L_{-1} +  G^{(2)}_{UV}e^{-\rho}W_{-1} + T_{UV}e^{-2\rho}W_{-2}\right)dx^-
\end{eqnarray}
to the gauge connection of  the interpolation solution \eqref{bsint}. Then solving the linearized equation of motion  leads to the relations between UV fields $\co_{UV}$ and IR fields $\co_{IR}$.

%%%%%%%%%%%%%%%%%%%%%%%%%
%%%%%%%%%%%%%%%%%%%%%%%%%
\section{Higher-spin supergravity with different $osp(1|2)$ embeddings}
%%%%%%%%%%%%%%%%%%%%%%%%%
%%%%%%%%%%%%%%%%%%%%%%%%%
In this section, we review the Chern-Simons formalism for higher-spin supergravity. We will use this formalism to study the spin-3 supergravity which can be expressed in terms of the Chern-Simons theory based on the Lie superalgebra $sl(3|2)$.  To prepare for later sections, we discuss the different $osp(1|2)$ embeddings of the $sl(3|2)$ algebra, which is the supersymmetric version of the $sl(2,\br)$ embedding of the $sl{(3,\br)}$ algebra.

%%%%%%%%%%%%%%%%%%%%%%%%%
\subsection{Higher-spin supergravity as a Chern-Simons theory}
%%%%%%%%%%%%%%%%%%%%%%%%%
Higher spin supergravity can be written as a Chern-Simons theory based on some superalgebra $\cg$. We consider the $\cn=2$ higher-spin supergravity, where the relevant gauge superalgebra is $\cg=sl(n|n-1)$  \cite{chr,hp}. For this choice the higher-spin gauge theory contains $\cn=2$ supermultiplets with spin ranging from 2 to $n$. The action reads
 \begin{equation} \label{action2}
I_{HS} = I_{CS}(\Gamma,k_{CS}) - I_{CS}(\tilde{\G},k_{CS})\;, \qquad
I_{CS} (\G,k_{CS})= \frac{k_{CS}}{4 \pi}  \int_{\cm} \mbox{STr}(\G \wedge d \G + \frac{2}{3} \G \wedge \G \wedge \G)\;,
\end{equation}
where $\G$ is the gauge super-connection that evaluates in $\cg$.  The connections can be expanded in terms of the bosonic and fermionic generators of $\cg$
\begin{equation}
\Gamma = {\displaystyle \sum_{s,m}} A^{(s)}_{m} L^{(s)}_{m} + {\displaystyle \sum_{s,r}}\psi^{(s)}_{r} G^{(s)}_{r}\;, \qquad
\tilde{\Gamma} ={\displaystyle \sum_{s,m}} \tilde{A}^{(s)}_{m} \tilde{L}^{(s)}_{m} + {\displaystyle \sum_{s,r}} \tilde{\psi}^{(s)}_{r} \tilde{G}^{(s)}_{r}\;,
\end{equation}
where $L^{(s)}_{m}~ (\tilde{L}^{(s)}_{m})$ are bosonic generators, $G^{(s)}_{m} ~(\tilde{G}^{(s)}_{m})$ are fermionic generators, $A^{(s)}_{m}~ (\tilde{A}^{(s)}_{m})$ are the bosonic fields and $\psi^{(s)}_{m} ~(\tilde{\psi}^{(s)}_{m})$ are fermionic fields.

We realize the superalgebra $\cg=sl(n|n-1)$ as supermatrices following the convention \cite{fss} where $\cg$ is spanned by matrices of the form
\begin{equation}\label{mr}
M = \left(\begin{array}{cc} E & B \cr C & D \end{array}\right)\,,
\end{equation}
in which $E$ and $D$ are $n\times n$ and $(n-1) \times (n-1)$ matrices that generate the $gl(n)$ and $gl(n-1)$ algebra respectively, $B$ and $C$ are
$n \times (n-1)$ and $(n-1) \times n$ matrices respectively. The ``STr'' in \eqref{action2} stands for the  super-trace and is defined as
\begin{equation}\label{str}
\mbox{STr}(M) =\mbox{Tr}(E) - \mbox{Tr}(D)\,.
\end{equation}
 Every element $g \in sl(n|n-1)$ satisfies $\mbox{STr}(g)=0$. Note that in this paper we follow the normalization of super-trace in \cite{fss}, which is different from that in \cite{hp}.

To minimize the confusion caused by different notations, we will use the representation of the $sl(3|2)$ algebra in the Racah basis following \cite{fl}. In \cite{fl}, the bosonic generators of $sl(n|n-1)$ are
\begin{equation}\label{bsrb}
T^{s}_{m}\,, \quad 0\leq s \leq n-1,~-s\leq m\leq s,  ~~~\text{and}~~~~ U^{t}_{u}\,, \quad\,~0 \leq t \leq n-2,~  ~-t\leq u\leq t\,,
\end{equation}
and the fermionic generators are
\begin{equation}\label{fmrb}
Q^{r}_{p}\,, \quad 1/2\leq r \leq n-3/2,\, -p\leq r\leq p, ~~~\text{and}~~~~\bar Q^{\bar r}_{q}\,,\quad 1/2\leq \bar{r} \leq n-3/2,\,-q\leq \bar r\leq q\,.
\end{equation}
Each of these generators is associated with a $(2n-1) \times (2n-1)$ matrix, see \cite{fl} for the concrete matrix realization of all the generators.

Coming back to \eqref{action2}, the relation between $k_{CS}$ and the parameter $\ell$ is similar to \eqref{kl}
\begin{equation}\label{skl}
k_{CS}=\frac{\ell}{4G_3}\frac{1}{(-{\mbox{STr}} (L^{(2)}_{0})^{2})}\,.
\end{equation}
where $L^{(2)}_{0,\pm1}$ are the bosonic generators of an $osp(1|2)$ subalgebra of $\cg$. The metric can be computed from the connection
\begin{equation}\label{sgenmetric}
g_{\m\n}=\frac{\ell^{2}}{4 \,\mbox{Tr} L_{0}^{2}}\mbox{STr} ((\tilde \G_{\m}- \G_{\m})(\tilde \G_{\n}- \G_{\n}))\,.
\end{equation}

%%%%%%%%%%%%%
\subsection{Normalization}\label{nmlz}
%%%%%%%%%%%%%%
In this paper, we will consider different $osp(1|2)$ subalgebras of $sl(n|n-1)$. Therefore, the ${\mbox{STr}} (L^{(2)}_{0})^{2})$ can be different for different embeddings. To compare results from different embeddings, we need a proper normalization.  To simplify our notation, we will omit the $(2)$ in $L^{(2)}_{0}$ throughout this paper.

As mentioned above, the key identity relating the frame-like language and the Chern-Simons language is \eqref{kl}. In the current supersymmetric case and following the definition \eqref{str} it becomes\begin{equation}\label{skeyeqn}
({\mbox{STr}} L_{0}^{2})\,k_{CS}=\frac{\ell}{4G_3}\,,
\end{equation}
where on the right hand side of the equation, the parameter $\ell$ is related to the cosmological constant and $G_{3}$ is the 3-dimensional gravity constant. On the left hand side of the equation, $k_{CS}$ is the Chern-Simons level and the ${\mbox{STr}} L_{0}^{2}$ is related to the different embeddings. This equation is valid for all different embeddings. Since the factor ${\mbox{STr}} L_{0}^{2}$ changes among different embeddings,  the $k_{CS}$ and $\ell$ cannot be fixed at the same time. We fix the value of $k_{CS}$ among different embeddings, since the higher spin action \eqref{action2} is unchanged when we identify different generators to span the $osp(1|2)$ subalgebra\fn{We thank Thomas Hartman for clarifying this point to us.}.

Then suppose we identify some different  $L'_{0,\pm1}$ as the bosonic generators of the $osp(1|2)$. We have  the identification:
\begin{equation}\label{kl2}
k_{CS}=\frac{\ell'}{4G_3}\frac{1}{({\mbox{STr}} (L'_{0})^{2})}\,.
\end{equation}
Since $k_{CS}$ is fixed in the two identifications, we develop a relation between $\ell$ and $\ell'$:
 \begin{equation}\label{2ls}
\ell/({\mbox{STr}} (L_{0})^{2})=\ell'/({\mbox{STr}} (L'_{0})^{2})\Rightarrow \L ({\mbox{STr}} (L_{0})^{2})^{2}=\L'({\mbox{STr}} (L'_{0})^{2})^{2}\,.
\end{equation}
Thus we see different identifications of the $osp(1|2)$ subalgebra, which in general have different values of ${\mbox{STr}} (L_{0})^{2}$, correspond to spacetimes with different cosmological constants.

%%%%%%%%%%%%%%%%%%%%%%
\subsection{$osp(1|2)$ embeddings of $sl(3|2)$}\label{ospemb}
%%%%%%%%%%%%%%%%%%%%%%
In this subsection, we construct the three different $osp(1|2)$ embeddings into the Lie superalgebra $sl(3|2)$. The results in this section are useful for later discussions.

In \cite{frs}, the authors discussed the $osp(1|2)$ embeddings of the $sl(n+1|n)$ superalgebra in general. For our case $sl(3|2)$, we will see that there are three different embeddings. Here we want to derive the explicit form of the decomposition of the adjoint representations of $sl(3|2)$ for each of the embeddings. In other words, we want to find the explicit grouping of the generators of the $sl(3|2)$ algebra into the representations of the $osp(1|2)$ subalgebra. This is a supersymmetric generalization of the bosonic $sl(2,\br)$ embedding considered in \cite{frs,agkp}. It is shown in \cite{frs} that any $osp(1|2)$ embedding in a basic Lie superalgebra $\cg$ can be considered as the superprincipal $osp(1|2)$ embedding of a regular
subsuperalgebra $\ck$ of $\cg$. Further notice that not all the basic Lie superalgebras admit an $osp(1|2)$ superprincipal embedding. The basic Lie superalgebras admitting a superprincipal $osp(1|2)$ are the following \cite{fss}:
\[
sl(n \pm 1|n),\,~ osp(2n \pm 1|2n),\,~ osp(2n|2n),\,~ osp(2n+2|2n),\,~
D(2,1;\alpha)~ \smbox{with} \alpha \ne 0, -1,\infty
\]
where $D(2,1;\alpha)$ is a one-parameter family of exceptional Lie superalgebras with rank 3 and dimension 17.
For the algebra $\cg=sl(3|2)$, there are three different regular subsuperalgebras $\ck=sl(3|2)$, $sl(2|1)$, $sl(1|2)$, so there are three different $osp(1|2)$ embeddings into $sl(3|2)$. We study each of the three embeddings in the following 3 subsections.

%%%%%%%%%%%%%%%%%%%%%%
\subsubsection{The principal embedding\,: $\ck=sl(3|2)$}\label{embprinp}
%%%%%%%%%%%%%%%%%%%%%%
This is the principal embedding. Following the notation in \cite{frs}, the decomposition of the adjoint representation of $sl(3|2)$, which is the algebra itself, reads
\be\label{principalembedding}
\frac{{\bf Ad}[sl(3|2)]}{sl(3|2)} =
\crr_2 \oplus \crr_{3/2} \oplus \crr_{1} \oplus \crr_{1/2} \,.
\ee
where $\crr_j $ is an irreducible representation of $osp(1|2)$ algebra with spin $j$. Notice that $\crr_j$ can be decomposed as $\crr_j=\cd_j \oplus \cd_{j-1/2}$  with $\cd_j$ being the spin-$j$ representation of $sl(2)$.
From this, we see that the bosonic sector of $sl(3|2)$,
\begin{equation}\label{bs32}
sl(3|2)\big|_{B}=sl(3)\oplus sl(2)\oplus
u(1)\,.
\end{equation}
decomposes as
\be\label{sl32bdcp21}
{\bf Ad}[sl(3|2)]\big|_{B} =
\cd_2 \oplus \cd_{1}\oplus \cd'_{1} \oplus \cd_0 \,.
\ee
Notice that $\cd_{2},\cd_{1},\cd_{0}$  comes from the $sl(3)$ subalgebra, while the $\cd'_{1}$ comes from the $sl(2)$ subalgebra.\fn{In this section, we use ``\,$'$\," to indicate any representation coming from the $sl(2)$ part in \eqref{bs32}.}
These correspond to a spin-$2$ representation,\fn{In this section, ``spin'' refers to the $sl(2)$ spin, the spacetime spin is the $sl(2)$ spin plus one.} two spin-$1$ representations and a spin-$0$ (singlet) representation, which agrees with \cite{hp}.

The fermionic sector contains
\be\label{sl32fdcp21}
{\bf Ad}[sl(3|2)]\big|_{F} =
2\cd_{3/2} \oplus 2 \cd_{1/2}\,.
\ee
We see there are 2 spin-$3/2$ representations and 2 spin-$1/2$ representations, which again agrees with the result in \cite{hp}.
As a simple consistency check, we count the number of bosonic and fermionic states. From \eqref{sl32bdcp21}, there are $5+3+3+1=12$ bosonic states.  From \eqref{sl32fdcp21}, there are $2\times 4+2 \times 2=12$ fermionic states, which is the same as the number of bosonic states. This agrees with the well known property of the $sl(n|n-1)$ supealgebra.

%%%%%%%%%%%%%%%%%%%%%%
\subsubsection{Non-principal embedding I\,: $\ck=sl(2|1)$}\label{non1}
%%%%%%%%%%%%%%%%%%%%%%
In this section, we consider the $osp(1|2)$ superprincipal embedding of $\ck= sl(2|1)$ in $\cg=sl(3|2)$
\be\label{sl32dcp10}
\frac{{\bf Ad}[sl(3|2)]}{sl(2|1)} =
\crr_1\oplus 3\crr_{1/2}  \oplus 2\tilde\crr_{1/2}  \oplus 2 \crr_0 \oplus 2 \tilde\crr_0\,,
\ee
where $ \tilde\crr_{j}=\tilde\cd_{j}\oplus\tilde\cd_{j-1/2}$ represents a spin-$j$ representation of $osp(1|2)$ but with  ``wrong'' statistics,  in the sense that $\tilde\cd_{i}$ is spanned by fermionic  (bosonic) generators if $i$ is an integer (half integer). In addition, $ \tilde\crr_{0}= \tilde\cd_{0}$, $\crr_{0}=\cd_{0}$. We see that there is only one $sl(2)$ subalgebra in  the decomposition \eqref{sl32dcp10}: the $\cd_{1} $ in $\crr_{1}$. The decomposition of the bosonic subalgebra \eqref{bs32} under this $osp(1|2)$ embedding reads
\be\label{sl32bdcp10}
{\bf Ad}[sl(3|2) ]\big|_{B}=
\cd_1 \oplus 2\tilde\cd_{1/2} \oplus 2 \cd_0 \oplus 3 \cd'_0\,.
\ee
Thus we see that the $sl(3)$ subalgebra in the decomposition \eqref{bs32} gives the unique $sl(2)$ subalgebra, 2 (bosonic) spin-$1/2$ representations of the $sl(2)$ and 2 singlets. The $sl(2)$  part in \eqref{bs32} decomposes into 3 singlets.

The decomposition of the fermionic part under this $osp(1|2)$ embedding reads
\be\label{sl32fdcp10}
{\bf Ad}[sl(3|2)]\big|_{F} =
4 \cd_{1/2} \oplus 4 \tilde\cd_{0}  \,.
\ee
We see the fermionic part breaks into 4 spin-$1/2$ representations  and 4 singlets of $sl(2)$. We can do a similar state counting here. From \eqref{sl32bdcp10}, there are $3+2\times 2+5\times 1=12$ bosonic states.  From \eqref{sl32fdcp10}, there are $4\times 2+4 \times 1=12$ fermionic states, which is the same as the number of bosonic states.

%%%%%%%%%%%%%%%%%%%%%%
\subsubsection{Non-principal embedding II\,: $\ck=sl(1|2)$}\label{non2}
%%%%%%%%%%%%%%%%%%%%%%
In this section, we consider the case  of $\ck=sl(1|2)$.
This embedding reads
\be\label{sl32bcp01}
\frac{{\bf Ad}[sl(3|2)]}{sl(1|2)} =\crr'_1\oplus 5 \crr_{1/2}\oplus 4\crr_0\,.
\ee
We can again decompose the bosonic subalgebra into spin-$s$ representations $\cd_s$ of $sl(2,\br)$
\be\label{sl32bdcp01}
{\bf Ad}[sl(3|2)] \big|_{B}=
\cd'_1 \oplus 9 \cd_0 \,.
\ee
Thus we learn that under this embedding, the bosonic subalgebra $sl(2)$ gives the only spin-$1$ representation, corresponding to the gravitational sector, and the  $sl(3)$ subalgebra decomposes to 9 singlets.

The decomposition of the fermionic part
in this $osp(1|2)$ embedding reads
\be\label{sl32fdcp01}
{\bf Ad}[sl(3|2)] \big|_{F}=
6 \cd_{1/2}\,.
\ee
 This means that  there are 6 spin-$1/2$ representations. From \eqref{sl32bdcp01}, there are $3+9\times 1=12$ bosonic states.  From \eqref{sl32fdcp01}, there are $6\times 2=12$ fermionic states, which agrees with the number of bosonic states.

%%%%%%%%%%%%%%%%%%%%%%%%%
%%%%%%%%%%%%%%%%%%%%%%
\section{The asymptotic symmetry}
%%%%%%%%%%%%%%%%%%%%%%
%%%%%%%%%%%%%%%%%%%%%%%%%
Following the procedure in \cite{cfpt,gh} for the bosonic case and \cite{hp} in the supersymmetric case, we can derive the asymptotic symmetries corresponding to the 3 different embeddings that we presented in section \ref{embprinp}, \ref{non1}, and \ref{non2}.

%%%%%%
\subsection{The asymptotic symmetry of the principal embedding}\label{emb1}
%%%%%%
We want to find the asymptotic symmetry algebra corresponding to the principal embedding. This can be done by a truncation of the result in \cite{hp}. However, for ease of comparison with the other embeddings, we compute it in a different representation, namely the explicit Racah basis realization \cite{fl}. This means all the elements of the Lie superalgebra $sl(3|2)$ are expressed in terms of bosonic bases \eqref{bsrb} and fermionic bases \eqref{fmrb}. As discussed in section \ref{ospemb}, for each $osp(1|2)$ embedding of the $sl(3|2)$ superalgebra, certain combinations of the $sl(3|2)$ generators form different  $osp(1|2)$ representations. We now construct these representations by explicitly giving their generators in terms of the Racah basis.

By definition \cite{frs,scheunert,dggt}, the following new generators characterize the decomposition of the $sl(3|2)$ algebra under the principal embedding
\begin{eqnarray}
\nn && \hspace{-1cm} L_0 = -\frac{U^1_0-2\,T^1_0}{\sqrt{2}}\,,\quad L_1= U^1_1+2\sqrt{2}\,T^1_1\,,\quad L_{-1}= U^1_{-1}+\sqrt{2}\,T^1_{-1}\,, \quad J=-2\sqrt{3} \, T^0_0-3\sqrt{2}\,U^0_0\\
\nn && \hspace{-1cm} K_0 = -\frac{\sqrt{2}}{3}(2\, U^1_0+\,T^1_0)\,,\qquad K_1= \frac{4}{3\sqrt{2}}(\sqrt{2} U^1_1+\,T^1_1)\,,\qquad K_{-1}= \frac{1}{3} (4 \,U^1_{-1}+\sqrt{2}\,T^1_{-1})\\
 && \hspace{-1cm} W_{2} = -4\, T^2_2\,, \quad W_{1} = \sqrt{2}\, T^2_1\,, \quad W_{0} = -\sqrt{\frac{2}{3}}\, T^2_0\,, \quad W_{-1} = \frac{T^2_{-1}}{\sqrt{2}}\,, \quad W_{-2} = - T^2_{-2} \label{prinpemb}
\end{eqnarray}
where $T^s_m\,,~U^t_u$ are the basis \eqref{bsrb}.
Note that this gives an explicit Racah basis realization of the truncated $shs[\l]$ algebra given in \cite{hp}.\fn{The connection to the result in \cite{hp} is through the map
$   J\to L^{(3/2)}_m\,,~ L_m\to L^{(2)}_m\,,~ K_m \to L^{(5/2)}_m\,,~ W_m\to L_{m}^{(3)}\,.$
}
In \eqref{prinpemb}, $L_i$ are the bosonic generators of the $osp(1|2)$ subalgebra, $K_i$ generate the other spin-1 representation, namely the $\cd'_1$ in \eqref{sl32bdcp21}. $W_i$ form the spin-2 representation and $J$ is a singlet.

 We spare the reader from the more complicated expressions of the fermionic generators, the form of them are determined so that the commutators of our principal embedding into the $sl(3|2)$ algebra agree with the truncated $shs[\l]$ commutators in \cite{hp} at $\l=-1$.

As shown in \cite{cfpt,hp}, after imposing the appropriate boundary conditions and gauge fixing, only the lowest mode\fn{Here we choose the so-called ``highest weight" or ``Drinfeld-Sokolov'' gauge. } of each spin-$s$ representation survives in the gauge connection $\G$. In our present case, we have
\begin{eqnarray}
\nn\G&=&\G_+dx^++\G_-dx^-+\G_\r d\r,\\[2mm]
\nn\G_+&=&e^{-\r L_0}\g e^{\r L_0},\qquad
\G_\r=e^{-\r L_0} \pa_\r e^{\r L_0},\qquad \G_-=0\,,\\
\nn\g&=&L_1+\frac{2 \p}{k_{CS}} (\cl L_{-1}+\cj J+\ca K_{-1}+\cw W_{-2}) \\
&&+\frac{2 \p}{k_{CS}}(\cg_{1/2}G^{1/2}_{-1/2}+\cg_{3/2}G^{3/2}_{-3/2}+\bar{\cg}_{1/2}\bar{G}^{1/2}_{-1/2}+\bar{\cg}_{3/2}\bar{G}^{3/2}_{-3/2}~)\,,\label{ggcnt0}
\end{eqnarray}
where all the calligraphic letters represent gauge fields that depend on $x^{+}$ only. For simplicity, we do not write out their $x^{+}$ dependence.

Now consider gauge transformations. As shown in \cite{cfpt}, the most general gauge transformation can be parameterized as $\L=\frac{k_{CS}}{2 \p}e^{-\r\hl_0} \l(x^+) e^{\r\hl_0}$, where
\begin{eqnarray}
\l&=&- \frac{1}{3}\sum\limits_{i=-1}^1\a_i L_i-\frac{1}{6}\r J+\frac{3}{4}\sum\limits_{i=-1}^1\b_i L_i+\frac{1}{4}\sum\limits_{i=-2}^2\g_i W_i\label{ggl0}\\
\nn&&-
\e^{1/2}_{-1/2} G^{1/2}_{-1/2}-\frac{1}{6}
\e^{1/2}_{1/2} G^{1/2}_{1/2}-
\e^{3/2}_{-3/2} G^{3/2}_{-3/2}-
\e^{3/2}_{-1/2} G^{3/2}_{-1/2}-
\e^{3/2}_{1/2} G^{3/2}_{1/2}+\frac{1}{4}
\e^{3/2}_{3/2} G^{3/2}_{3/2}\\
\nn&&-\bar{\e}^{1/2}_{-1/2}  \bar{G}^{1/2}_{-1/2}+\frac{1}{6}
\bar{\e}^{1/2}_{1/2}  \bar{G}^{1/2}_{1/2}-
\bar{\e}^{3/2}_{-3/2}  \bar{G}^{3/2}_{-3/2}-
\bar{\e}^{3/2}_{-1/2}  \bar{G}^{3/2}_{-1/2}-
\bar{\e}^{3/2}_{1/2}  \bar{G}^{3/2}_{1/2}-\frac{1}{4}
\bar{\e}^{3/2}_{3/2} \bar{G}^{3/2}_{3/2}
\end{eqnarray}
and the Greek letters $\a,\b,\g,\r,\e,\bar{\e}$ stand for the gauge parameters associated with different generators. Their dependence on the $x^+$ is suppressed for simplicity. All different numerical factors are present to make sure that the global charge has the following form\fn{Note that we take a different approach here comparing to the computation in \cite{gh, hp} where the numerical factors are multiplied to the components in $\g$ ($N_{i}$ in \cite{gh} and $N_{i}^{\text{B(F)}}$ in \cite{hp}). We take this approach in this computation to be consistent with the computations of  the two non-principal embeddings in section \ref{nonem1} and \ref{emb2}, where the super-trace structures of the generators in the non-principal embeddings are complicated. This point will be explained later. }
\begin{eqnarray}
 \hspace{-10mm} Q(\L)\hspace{-3mm}&=&\hspace{-3mm}\int d\f\, \mbox{STr}(\G(x^+)\L(x^+))\nn\\
  \hspace{-3mm}&=&\hspace{-4mm} \int d\f \bigg(\a_1\cl+\r\cj +\b_1 \ca +\g_2\cw W_{-2}+\e^{1/2}_{1/2}\cg_{1/2}+\e^{3/2}_{3/2}\cg_{3/2}+\bar{\e}^{1/2}_{1/2}\bar{\cg}_{1/2}+\bar{\e}^{3/2}_{3/2}\bar{\cg}_{3/2}\bigg)\,,\label{generalcharge0}
\end{eqnarray}
where the implicit $x^+$ dependence of the integrand is understood. The integration variable $\f$ is the angular coordinate of AdS$_{3}$ and is related to the light-cone coordinates by $x_{\pm}=t/\ell\pm \f$. The integral \eqref{generalcharge0} is evaluated at constant time so the gauge fields are $\f$ dependent only, which makes sense of the integration \eqref{generalcharge0}.

Now we proceed to compute the asymptotic symmetry algebra. We follow the exact procedure of \cite{gh, hp}. The first step is to compute the variation of the gauge fields \eqref{ggcnt0} under the gauge transformation with parameter $\l$ \eqref{ggl0}. The variation is given by
\begin{equation}\label{npcharge}
 \del_\l \g=d\l+[\g,\l]\,.
\end{equation}
Plugging \eqref{ggcnt0} and \eqref{ggl0} into \eqref{npcharge} gives the variation of the various fields in the connection \eqref{ggcnt0}.
The asymptotic symmetry is defined to be the gauge transformations, parameterized by $\l$, that preserve the form of \eqref{ggcnt0}. This means only the lowest mode of each spin-$s$ representation appears in $\del_{\l} \g$ \eqref{npcharge}, which leads to constraints on the gauge parameters. To find the Poisson bracket we apply the following relation\fn{We follow the convention of \cite{hms} for consistency.}
\begin{equation}\label{vrn2pb0}
 \del_\l V=\{Q(\l), V \}\,.
\end{equation}
Combining \eqref{generalcharge0}, \eqref{npcharge} and \eqref{vrn2pb0} and imposing the constraints on the gauge parameters, we get the expression for the Poisson brackets.

To get the standard form of the commutators of the super-Virasoro algebra, we need a rescaling of the stress tensor\fn{Note that this rescaling is a consequence of putting the super-trace factors in the gauge parameter \eqref{ggl0} in our approach. In \cite{hp}, no rescaling is needed since we use a different approach. }
\begin{equation}
  \cl \rightarrow 3\cl\,,
\end{equation}
and a shift of the zero mode
\begin{equation}
  \cl \rightarrow \cl-\frac{k_{CS}}{4 \p}\,.
\end{equation}

A further Sugawara type modification to the stress-tensor $\cl$
\begin{equation}\label{sug0}
  \cl\to\cl+\frac{\p}{6 k_{CS}}\cj^2\,,
\end{equation}
is also performed to ensure that the current $\cj$ has the correct conformal dimension $\Delta=1$.

To express the algebra in the commutators among modes, we adopt the conversion from the Poisson brackets to the Dirac brackets by $i\{\cdot,\cdot\}_\text{PB}=[\cdot,\cdot]$ \cite{hms}. Then the following mode expansion
\begin{eqnarray}
&&\cl(\f)=\frac{1}{2\p}\sum\limits_{n} L_{n} e^{in\f}
\end{eqnarray}
gives the commutation relation
\begin{eqnarray}
&&[ L_{m}, L_{n}] = (m-n) L_{m+n}+\frac{c}{12}(m^{3}-m)\delta_{m,-n}\,,
\end{eqnarray}
where the central charge is read off as $c=18 k_{CS}$. This central charge is different from the central charge in \cite{hp} due to the different super-trace normalization $\mbox{STr}L_{0}^{2}=\frac{3}{2}$ in this paper. If we had chosen the normalization used in \cite{hp}, namely $\mbox{STr}'L_{0}^{2}=\frac{1}{2}$, the resulting central charge will be the same as that in \cite{hp}. So the computation here really reproduces the result in \cite{hp}. In a similar fashion, we can work out the commutation relations among other fields, but since the result is the same as in \cite{hp} after a truncation by setting $\l=-1$, we omit the other commutators. For the readers who are interested in the full result, section 3 in \cite{ hp} will be useful.

The important lessons from this section are (i) the asymptotic symmetry corresponding to the principal embedding being the super-$\cw_{3}$ algebra \cite{ito93} and (ii) the value of the central charge being $c=18 k_{CS}$.

%%%%%%
\subsection{The asymptotic symmetry of the non-principal  embedding I}\label{nonem1}
%%%%%%

Motivated by \eqref{sl32dcp10}, we realize the non-principal embedding in section \ref{non1} by giving explicit expressions for the generators of the different $osp(1|2)$ representations in \eqref{sl32dcp10}. In terms of  the Racah basis, the bosonic generators read
\begin{subequations}\label{othebd1b}
\begin{equation}\label{spin2hat}
\hl_{0}=-T^{1}_{0}/\sqrt{2},\qquad \hl_{1}=T^{2}_{2},\qquad \hl_{-1}=-T^{2}_{-2}\,,
\end{equation}
\vspace{-5mm}
\begin{equation}
\hm_{1/2}= (T^{1}_{1}-T^{2}_{1})/(4\,x),\qquad \hm_{-1/2}= (T^{1}_{-1}+T^{2}_{-1})/(4\,x)\,,
\end{equation}
\begin{equation}\label{spin32hat}
\hn_{1/2}= (T^{2}_{1}+T^{1}_{1}) x,\qquad \hn_{-1/2}= (T^{2}_{-1}-T^{1}_{-1}) x\,,
\end{equation}
\vspace{-3mm}
\begin{equation}\label{spin1hat}
\hj_{1}=\frac{i}{6}\left(2 \sqrt{3} T^{0}_{0}+\sqrt{2} \left(\sqrt{3} T^{2}_{0}+3  U^{0}_{0}+3 U^{1}_{0})\right)\right),
\end{equation}
\begin{equation}
\hj_{2}=-i U^{1}_{-1},\qquad
\hj_{3}= i U^{1}_{1},
\end{equation}
\begin{equation}
\hj_{4}=\frac{i}{6}\left(2 \sqrt{3} T^{0}_{0}+\sqrt{2} \left(\sqrt{3} T^{2}_{0}+3 ( U^{0}_{0}- U^{1}_{0})\right)\right),
\end{equation}
\begin{equation}
\hj_{5}=i\sqrt{3/2}~T^{2}_{0}\,.
\end{equation}
\end{subequations}
The bosonic parameter $x$ can be any nonzero constant and is largely irrelevant. The $T^{s}_{m}, U^{t}_{n}$ are the supersymmetric Racah basis. The various ``$i$" in the definition of $\hj_{m}$ are not necessary, they are introduced so that the resulting asymptotic symmetry algebra manifestly matches with the known result in \cite{dth}. We could just as well have defined a set of real bases, but  then we would have to do some further field redefinitions after we find the asymptotic algebra in order to match it with the known result.

The $\{\hl_{-1},\hl_{0},\hl_{1}\}$ generate the $sl(2)$ subalgebra. The  $\{\hm_{-1/2},\hm_{1/2}\}$ and $\{\hn_{-1/2},\hn_{1/2}\}$ form 2 bosonic spin-$1/2$ representations of the $sl(2)$ and the $\hj_{i},\,i=1,..., 5$ are the 5 $sl(2)$ singlets.

We also define the following 12 fermionic fields
\begin{subequations}
\vspace{-3mm}
\begin{equation}
\hg^{-1}_{1/2}=-\sqrt{2}{Q}^{3/2}_{3/2}/(2 \sqrt{2}~ x),  \qquad \hg^{-1}_{-1/2}=(\sqrt{2} {Q}^{3/2}_{-1/2}+2{Q}^{1/2}_{-1/2})/ (4\sqrt{3}~x)\,,\label{othebd1f}
\end{equation}
\vspace{-3mm}
\begin{equation}
\hg^{-2}_{1/2}=-(2{Q}^{1/2}_{1/2}-\sqrt{2} {Q}^{3/2}_{1/2})/ (4\sqrt{3}~x),\qquad \hg^{-2}_{-1/2}=- {Q}^{3/2}_{-3/2}/(2 \sqrt{2} x)\,,
\end{equation}
\vspace{-3mm}
\begin{equation}
\hg^{1}_{1/2}=x~(\sqrt{2} \bar{Q}^{3/2}_{1/2}+2\bar{Q}^{1/2}_{1/2})/ \sqrt{3},\qquad \hg^{1}_{-1/2}=\sqrt{2}~x~  \bar{Q}^{3/2}_{-3/2}\,,
\end{equation}
\begin{equation}
\hg^{2}_{1/2}=\sqrt{2}~x ~ \bar{Q}^{3/2}_{3/2},\qquad \hg^{2}_{-1/2}=x~(\sqrt{2} \bar{Q}^{3/2}_{-1/2}-2\bar{Q}^{1/2}_{-1/2})/ \sqrt{3}\,,
\end{equation}
\begin{equation}\label{spin1hatf}
\hx_{1}=i(\sqrt{2} \bar{Q}^{3/2}_{-1/2}+\bar{Q}^{1/2}_{-1/2})/ \sqrt{3},\qquad \hx_{2}=-i(\sqrt{2} {Q}^{3/2}_{1/2}+{Q}^{1/2}_{1/2})/ \sqrt{3}\,,
\end{equation}
\vspace{-3mm}
\begin{equation}
\hx_{3}=-i(\bar{Q}^{1/2}_{1/2}-\sqrt{2} \bar{Q}^{3/2}_{1/2})/ \sqrt{3},\qquad \hx_{4}=-i({Q}^{1/2}_{-1/2}-\sqrt{2} {Q}^{3/2}_{-1/2})/ \sqrt{3}\,.
\end{equation}
\end{subequations}

Now we can perform the standard procedure to get the asymptotic symmetry algebra. Similar to the computation of the principal embedding in the previous subsection, we have:
\begin{eqnarray}
\nn\G&=&\G_+dx^++\G_-dx^-+\G_\r d\r,\\[2mm]
\nn\G_+&=&e^{-\r\hl_0}\,\g\, e^{\r\hl_0},\qquad
\G_\r=e^{-\r\hl_0} \,\pa_\r e^{\r\hl_0},\qquad \G_-=0\,,\\
\nn\g&=&\hl_1+\frac{2 \p}{k_{CS}} \cl \hl_{-1}+\frac{2 \p}{k_{CS}} (\cm \hm_{-1/2}+\cn \hn_{-1/2}+\sum\limits_{i=1}^5 \cj_i \hj_i)\\
&&+\frac{2 \p}{k_{CS}}(\hspace{-3mm}\sum\limits_{~~~i=-1,-2,1,2}\hspace{-5mm}\cg_i\hg^i_{-1/2}+\sum\limits_{i=1}^4\cx_i \hx_i~)\,,\label{ggcnt}
\end{eqnarray}
where all the fields, denoted by calligraphic letters, only depend on $x^+$.

Now consider gauge transformations. The most general gauge transformation can be parameterized as $\L=\frac{k_{CS}}{2 \p}e^{-\r\hl_0} \l(x^+) e^{\r\hl_0}$, where
\begin{eqnarray}
\nn\l&=&- \sum\limits_{i=1}^3\a_i \hl_i+2\sum\limits_{i=\pm1/2}\n_i \hm_i-2 \sum\limits_{j=\pm1/2}\m_j \hn_j\\
&&\hspace{-3mm}+
(2 \r_1+\r_4-\r_5) \hj_1+ \r_3 \hj_2+ \r_2 \hj_3+ (\r_1+ 2\r_4- \r_5) \hj_4-( \r_1+ \r_4) \hj_5
\label{gaugept2}\\
[2mm]
\nn&&\hspace{-3mm}+\hspace{-3mm}\sum\limits_{~~~i=-1,-2,1,2}\hspace{-5mm}\s_i\hg^{i}_{-1/2}~-\hspace{-5mm}\sum\limits_{~~~i=-1,-2,1,2}\hspace{-5mm}2\,\e_i\hg^{-i}_{-1/2}
-\h_1\hx_2+\h_2\hx_1-\h_3\hx_4+\h_4\hx_3\,.
\end{eqnarray}
and the Greek letters $\a,\m,\n,\r,\s,\e,\h$ stand for the gauge parameters again. We suppress their $x^+$ dependence for simplicity.

The form of the gauge parameters \eqref{gaugept2} looks strange, especially the $\r_{i}$ parameters. This is due to the super-trace structure of the non-principal embedding. The super-trace has the following general property
\begin{equation}\label{generalcharge2}
  \mbox{STr}(L^s_m L^t_{-m})\propto \del^{st}\,.
\end{equation}
This means that a product of generators with different spins (i.e.~$s\neq t$) has vanishing super-trace. However, in this non-principal embedding, there are multiple fields with spin $s=1$, so the delta function in \eqref{generalcharge2} does not exclude mixings between different spin-$1$ generators. To make our derivation of the asymptotic symmetry algebra straightforward, we choose the form of the gauge parameters such that the global charge, defined as
\begin{equation}\label{generalcharge}
  Q(\L)=\int d\f\, \mbox{STr}(\G(x^+)\L(x^+))\,,
\end{equation}
have the indicative form
\begin{equation}\label{npcharge1}
 Q(\g)= \int d \f \, \bigg(\a_1 \cl+\m_{1/2} \cm+\n_{1/2} \cn+\sum\limits_{i=1}^5 \r_i \cj_i+\sum\limits_i^4  \e_i\cg_i+\sum\limits_i^4\h_i \cx_i\bigg) \,,
\end{equation}
where the relations between the fields, labelled by calligraphic letters, and the gauge transformation parameters, labelled by Greek letters, are clear. This simple form of the global charge \eqref{npcharge1} and the mixing between spin-$1$ generators are the origin of the strange combinations of the $\r_{i}$ parameters in \eqref{gaugept2}. If we had chosen a simple expression for the gauge parameters as in the principal embedding \eqref{ggl0}, the extraction of the asymptotic algebra is still possible but highly involved.

Now we proceed to compute the asymptotic algebra.
Following the same procedure as in the principal embedding, we get the expression for the Poisson brackets. A Sugawara type modification to the stress-tensor $\cl$
\begin{equation}\label{sug2}
  \cl\to\cl+\frac{\p}{2 k_{CS}}(\cj_1^2+\cj_4^2+3\cj_5^2+2\cj_1\cj_4+2\cj_1\cj_5+2\cj_4\cj_5+4\cj_2\cj_3+4\cx_2\cx_1+4\cx_4\cx_3)
\end{equation}
and a shift of the zero mode
\begin{equation}
  \cl \rightarrow -(\cl-\frac{k_{CS}}{4 \p})\,.
\end{equation}
are performed to put the commutators in the form of the standard super-Virasoro algebra. Recall that when we convert from the Poisson brackets to the Dirac brackets, we use the relation $i\{\cdot,\cdot\}_\text{PB}=[\cdot,\cdot]$. The factor ``$i$" in this relation is part of the reason why we complexify the generators of the $sl(2)$ singlets $\hj_i$ \eqref{othebd1b} and $\hx_i$ \eqref{othebd1f}. Finally, the following mode expansions
\begin{eqnarray}
\nn &&\cl(\f)=\frac{1}{2\p}\sum\limits_{n} L_{n} e^{in\f}\,,\qquad  \cj_{a}(\f)=\frac{1}{2\p}\sum\limits_{n} J^{a}_{n} e^{in\f}\,,\qquad  \cx_{a}(\f)=\frac{1}{2\p}\sum\limits_{n} X^{a}_{n} e^{in\f}\,,\\
\nn &&\cg_{j}(\f)=\frac{1}{2\p}\sum\limits_{n} G^{j}_{n} e^{in\f }\,, \qquad \bar{\cg}_{j}(\f)=\frac{1}{2\p}\sum\limits_{n} \bar{G}^{j}_{n} e^{in\f}\,,\\
\nn && \cm_{j}(\f)=\frac{1}{2\p}\sum\limits_{n} M^{j}_{n} e^{in\f}\,, \qquad \cn_{j}(\f)=\frac{1}{2\p}\sum\limits_{n} N^{j}_{n} e^{in\f}\,.
\end{eqnarray}
give us the commutation relations
\begin{subequations}\label{u21alg}
\begin{eqnarray}
&&[ L_{m}, L_{n}] = (m-n) L_{m+n}+\frac{c}{12}(m^{3}-m)\delta_{m,-n}\,,  \\
&&[ L_{m}, J^{a}_{n}] = -n J^a_{m+n}\,,\\
&&[L_{m},G^{i}_{n}]=(\frac{m}{2}-n) G_{m+n}\,,\\
&&[L_{m},\bar{G}^{i}_{n}]=(\frac{m}{2}-n) \bar{G}_{m+n}\,,\\
&&[L_{m},M_{n}]=(\frac{m}{2}-n) M_{m+n}\,,\\
&&[L_{m},N_{n}]=(\frac{m}{2}-n) N_{m+n}\,,\\
&&[J^{a}_{m}, J^{b}_{n}] =f^{abc}\,J^{c}+ m\, k_{CS}\, h^{ab}\,\del_{m+n}\,,\label{jj1}\\
&&[J^{a}_{m}, X^{b}_{n}] =g^{abc}\,X^{c}\,,\\
&&\{X^{p}_{m}, X^{q}_{n}\} =\g^{pqc}\,J^{c}+m\, k_{CS}\, \b^{pq}\,\del_{m+n}\,,\label{xx1}\\
&&[J^{a}_{m}, G^{i}_{n}] =\q^{aij} G^{j}_{m+n}\,,\\
&&[J^{a}_{m}, \bar{G}^{i}_{n}] =\bar{\q}^{aij} \bar{G}^{j}_{m+n}\,,\\
&&[J^{5}_{m}, N_{n}] =N_{m+n}\,,\\
&&[J^{5}_{m}, M_{n}] =-M_{m+n}\,,\\
&&\{X^{2}_{m}, G^{1}_{n}\} =N_{m+n}\,, ~~ \qquad\qquad\qquad \{X^{4}_{m}, G^{2}_{n}\} =N_{m+n}\,,\\
&&\{X^{1}_{m}, \bar{G}^{1}_{n}\} =\{X^{3}_{m}, \bar{G}^{2}_{n}\} =-M_{m+n}\,,\\
&&[X^{1}_{m}, N_{n}] = G^{1}_{m+n}\,, ~\quad\qquad\qquad\qquad [X^{3}_{m}, N_{n}] = G^{2}_{m+n}\,,\\
&&[X^{2}_{m}, M_{n}] = \bar{G}^{1}_{m+n}\,, \,\quad\qquad\qquad\qquad [X^{4}_{m}, M_{n}] = \bar{G}^{2}_{m+n}\,,\\
&&[\cn_{m},\bar{G}^{i}_{n}]=-2 (m-n) \cx_{2i}, \qquad\qquad [\cm_{m},G^{i}_{n}]=2 (m-n) \cx_{2i-1}\\
\nn&&[M_m,N_n]= 2 L_{m+n}-(m-n)(J^1_{m+n}+J^4_{m+n}+3J^5_{m+n})\\
&&\qquad\qquad\qquad+U_{ab}^{33} (J^aJ^b)_{m+n}+V_{ab}^{33} (X^aX^b)_{m+n}+2 k_{CS}(m^2-\frac{1}{4})\del_{m,-n}\,,\\
\nn&&\{S^i_m,\bar{S}^j_n\}=-2L_{m+n}\del^{ij}+(m-n)[\del^{ij}(J^1_{m+n}+J^4_{m+n}+J^5_{m+n})-2J^{2(i-1)+j}]\\
 &&\qquad\qquad\qquad+U_{ab}^{ij} (J^aJ^b)_{m+n}+V_{ab}^{ij} (X^aX^b)_{m+n}-2 k_{CS}(m^2-\frac{1}{4})\del^{ij}\del_{m,-n}\,,
\end{eqnarray}
\end{subequations}
where $a,b=1,\ldots, 5\,, ~~p,q=1,\ldots, 4\,, ~~ i,j=1,\ldots, 2$. The central charge is read off from the algebra as $c=6 k_{CS}$. The non-vanishing structure constants $f^{abc},\,g^{abc},\,\q^{aij},\,\bar\q^{abc},\,\g^{pqc}$ are listed in Appendix \ref{strcnt}. The expressions for $U_{ab}^{ij}$ and  $V_{ab}^{ij}$ are complicated and are not enlightening since they will receive quantum corrections from the normal ordering, therefore we omit the expressions for them.

This algebra \eqref{u21alg} is the non-linear $u(2|1)$-extended superconformal algebra discovered in \cite{dth}. To match our expressions \eqref{u21alg} with the expressions in \cite{dth}, we employ the following dictionary
\begin{eqnarray}
 \nn && \qquad \cj_{1} \rightarrow  \S^{11},\qquad \cj_{2} \rightarrow  \S^{12},\qquad \cj_{3} \rightarrow  \S^{21},\qquad \cj_{4} \rightarrow  \S^{22},~~ \\
\nn  &&\cj_{5} \rightarrow  \S^{33}, \qquad \cx_1 \rightarrow \S^{13},\qquad
    \cx_2 \rightarrow \S^{31},\qquad
    \cx_3 \rightarrow \S^{23},\qquad
    \cx_4 \rightarrow \S^{32}\,,\\
    && \cg_i \rightarrow G^i, \quad i=1,2\,, \qquad \cg_i \rightarrow \bar{G}^{-i}, \quad i=-1,-2\,,\quad \cn \rightarrow G^3\,,\quad\cm \rightarrow \bar{G}^3\,,
\end{eqnarray}
at the following value of the parameters in \cite{dth}
\begin{eqnarray}
\nn  &&\k = k_{CS}\,,\qquad \k'=  k_{CS} \,,\qquad \e=0\,,\qquad\f=-2\,,\qquad \f'=-1\,,\qquad a=-2k_{CS}\,.
\end{eqnarray}
As stated in \cite{dth}, the resulting algebra is a nonlinear superconformal algebra (SCA) with $u(2|1)$-supersymmetry. We call this algebra $u(2|1)$-SCA. The spin-1 currents $\cj_{1\ldots 5}, ~\cx_{i=1,4}$ fall into the adjoint representation of the Lie superalgebra $u(2|1)$. The multiplet $(\cg_{1},\,\cg_{2},\,\cn)$ and $(\cg_{-1},\,\cg_{-2},\,\cm)$ are the fundamental and anti-fundamental representations of the $u(2|1)$ respectively. The spin-2 field $\cl$ is a singlet of $u(2|1)$.

$\bullet$ {\bf The quantum algebra.}
Note that what we have derived here is the classical version of the $\cn=2$ super-conformal algebra with quadratic non-linearity \cite{dth}. To get the full quantum algebra, we require all the commutators to satisfy the Jacobi identities. We can start with our classical result, set all coefficients of the non-linear commutators as well as the central charge
as independent parameters. Requiring this ansatz to satisfy the Jacobi identities yields the full quantum algebra. The result turns out to be the same as those presented in \cite{dth}, which is not surprising since our classical solution coincides with the ansatz in  \cite{dth}.  The key result for us is the new quantum corrected central charge
\begin{equation}
c=6k_{CS}+3\,.
\end{equation}
For the full algebra that satisfying the Jacobi identities, we refer the reader to \cite{dth}.

%%%%%%
\subsection{The asymptotic symmetry of the non-principal embedding II}\label{emb2}
%%%%%%
 The non-principal embedding \eqref{sl32bcp01}  implies the following realization for the generators of the different $osp(1|2)$ representations. In the Racah basis, the bosonic generators read
\begin{subequations}\label{othebd2}
\begin{equation}\label{spin2hat2}
\hhl_{0}=-U^{1}_{0}/\sqrt{2},\qquad \hhl_{1}=U^{1}_{1},\qquad \hhl_{-1}=U^{1}_{-1}\,,
\end{equation}
\vspace{-5mm}
\begin{eqnarray}
&& \hhj_1=\frac{1}{2\sqrt{2}}(T^1_{-1}-T^1_{1}+T^2_{-1}-T^2_{1})\,,\qquad
\hhj_2=\frac{-i}{2\sqrt{2}}  (T^1_{-1}+T^1_{1}+T^2_{-1}+T^2_{1})\,,\qquad\\
&&\hhj_3=\frac{-1}{2\sqrt{2}} \left(T^1_{0}+\sqrt{3} T^2_{0}\right)\,,\qquad
\hhj_4=\frac{T^2_{-2}+T^2_{2}}{{2}}\,,\qquad
\hhj_5=-\frac{i (T^2_{-2}-T^2_{2})}{{2}}\,,\qquad\\
&&\hhj_6=\frac{1}{2\sqrt{2}} (-T^1_{-1}+T^1_{1}+T^2_{-1}-T^2_{1})\,,\qquad
\hhj_7=\frac{i}{2\sqrt{2}} (T^1_{-1}+T^1_{1}-T^2_{-1}-T^2_{1})\,,\qquad\\
&&\hhj_8=\frac{1}{2\sqrt{2}} \left(-\sqrt{3} T^1_{0}+T^2_{0}\right)\,,\qquad
\hhj_9={-i} \left({T^0_{0}}+\sqrt{\frac{3}{2}} U^0_0\right)\,.
\end{eqnarray}
\end{subequations}
Note again that the various ``$i$'' are not necessary, they are introduced for the ease of comparison with known results, we can define the basis without complexification equally well.

The fermionic generators read
\begin{subequations}
\begin{eqnarray}
&&\hhg^{-1}_{-1/2}=-\frac{1}{2}Q^{{3/2}}_{-{3/2}}\,,
\qquad
\hhg^{-1}_{1/2}=\frac{1}{\sqrt{6}} Q^{{1/2}}_{{1/2}}+\frac{Q^{{3/2}}_{-{1/2}}}{2\sqrt{3}}\,,\\
&& \hhg^{-2}_{-1/2}=\frac{Q^{{1/2}}_{-{1/2}}}{2\sqrt{3}}-\frac{1}{\sqrt{6}} Q^{{3/2}}_{-{1/2}}\,,
\qquad
\hhg^{-2}_{1/2}=\frac{Q^{{1/2}}_{{1/2}}}{2\sqrt{3}}+\frac{1}{\sqrt{6}}Q^{{3/2}}_{{1/2}}\,,\\
&&\hhg^{-3}_{-1/2}=\frac{1}{\sqrt{6}} Q^{{1/2}}_{{1/2}}+\frac{Q^{{3/2}}_{{1/2}}}{2\sqrt{3}}\,,
\qquad
\hhg^{-3}_{1/2}=-\frac{1}{2}Q^{{3/2}}_{{3/2}}\,,\\
&&\hhg^{1}_{-1/2}=-\sqrt{\frac{2}{3}} \bar{Q}^{{1/2}}_{{1/2}}-\frac{\bar{Q}^{{3/2}}_{{1/2}}}{\sqrt{3}}\,,
\qquad
\hhg^{1}_{1/2}=\bar{Q}^{{3/2}}_{{3/2}}\,,\\
&&\hhg^{2}_{-1/2}=\frac{\bar{Q}^{{1/2}}_{-{1/2}}}{\sqrt{3}}+\sqrt{\frac{2}{3}} \bar{Q}^{{3/2}}_{-{1/2}}\,,
\qquad
\hhg^{2}_{1/2}=\frac{\bar{Q}^{{1/2}}_{{1/2}}}{\sqrt{3}}-\sqrt{\frac{2}{3}} \bar{Q}^{{3/2}}_{{1/2}}\,,\\
&&\hhg^{3}_{-1/2}= \bar{Q}^{{3/2}}_{-{3/2}}\,,
\qquad
\hhg^{3}_{1/2}=\sqrt{\frac{2}{3}} \bar{Q}^{{1/2}}_{-{1/2}}-\frac{\bar{Q}^{{3/2}}_{-{1/2}}}{\sqrt{3}}\,.
 \end{eqnarray}
\end{subequations}

One difference of this embedding is that the super-trace $\mbox{STr}L_0^2 <0$. Then from the relation \eqref{skeyeqn}, keeping $k_{CS}$ unchanged means $\ell$ changes sign. But this is fine: $\ell$ itself is not so important since it is related to the cosmological constant by $\L=-\frac{2}{\ell^{2}}$, so it is $|\ell|$ that represents the AdS radius.

A more convenient way to understand this extra minus sign goes like this. After we get the embedding \eqref{sl32bcp01}, we turn off all other fields and leave only the gravitational sector. The resulting action of the higher-spin theory\fn{The superscript ``$(2)$" represents the fact that only the spacetime spin-2 sector is left after the truncation.}
\begin{equation}\label{trunchs}
  I_{HS}^{(2)}=I^{(2)}_{CS}(\G,k_{CS})-I^{(2)}_{CS}(\tilde\G,k_{CS})\,,
\end{equation}
is the same as the Einstein-Hilbert action \eqref{action1} with a similar identification as \eqref{gaugefields}. However, as discussed above, $\ell<0$ for this non-principal embedding. This means the relation between the gauge fields and the vielbeins are
\begin{equation}\label{newviel}
  \G^{(2)}=\o^{(2)}-\frac{1}{|\ell|}e^{(2)}\,, \qquad \tilde\G^{(2)}=\o^{(2)}+\frac{1}{|\ell|}e^{(2)}\,.
\end{equation}
This looks a little unfamiliar comparing with the more conventional \eqref{gaugefields} where $\ell>0$ by default and is interpreted as the AdS radius. Then it is natural to consider the following rewriting of the action \eqref{trunchs}
\begin{equation}\label{newactiontr}
   I_{HS}^{(2)}=I^{(2)}_{CS}(\G',-k_{CS})-I^{(2)}_{CS}(\tilde\G',-k_{CS})\,,
\end{equation}
where $\G'=\tilde\G$ and $\tilde\G'=\G$. This is identical to the original formula \eqref{trunchs} and we simply switch the role of the $\G$ and $\tilde\G$. The benefit of this rewriting is that the relations between the new gauge connection $\G^{'(2)} ~(\tilde\G^{'(2)})$ and the vielbein/spin-connection take the conventional form as \eqref{gaugefields}
\begin{equation}\label{newfdviel}
  \G^{'(2)}=\o^{(2)}+\frac{1}{|\ell|}e^{(2)}\,, \qquad \tilde\G^{'(2)}=\o^{(2)}-\frac{1}{|\ell|}e^{(2)}\,.
\end{equation}
After this redefinition, nothing is special for this embedding except for the minus sign in front of the $k_{CS}$ in \eqref{newactiontr}. This minus sign is crucial to get the correct expression for the central charge in the later computation.

This rewriting \eqref{newactiontr} should be carried over to the full higher-spin gauge connections for consistency and we will use the relabeled gauge connections $\G'$ and $\tilde\G'$ to do all the following computations. But to simpify the notation, we drop the ``$\,'\,$" everywhere. Then the computation can be carried out in exactly the same was as in the previous two subsections with only one important thing to keep in mind: the Chern-Simons level used in this computation becomes $-k_{CS}$. Now we carry out the explicit computation. The super-connection takes the general form as in \eqref{ggcnt} after imposing the asymptotic AdS boundary condition gauge fixing
 \begin{eqnarray}
\nn\G&=&\G_+dx^++\G_-dx^-+\G_\r d\r,\\[2mm]
\nn\G_+&=&e^{-\r\hl_0}\,{\g}\, e^{\r\hl_0},\qquad
\G_\r=e^{-\r\hl_0} \,\pa_\r e^{\r\hl_0},\qquad \G_-=0\,,\\
\nn{\g}&=&\hl_{-1}+\frac{2 \p}{-k_{CS}} \bigg(\cl \hl_{1} +\sum\limits_{i=1}^9 \cj_i\, \hj_i+\hspace{-3mm}\sum\limits_{~~~i=-1,-2,-3,1,2,3}\hspace{-5mm}\cg_i\,\hg^i_{1/2}\bigg)\,.\label{ggcnt2}
\end{eqnarray}

In this case, the gauge parameter $\L$ takes the  form
\begin{eqnarray}
\nn\hat{\hat{\L}}&=&\frac{-k_{CS}}{2 \p}e^{-\r\hl_0} \hhlm(x^+) e^{\r\hl_0}\,,\\
\hhlm&=&\hspace{-3mm}-\bigg(2 \sum\limits_{i=1}^8 \hat{\r}_i \hhj_i+\hat{\r}_9 \hhj_9 (6 h^{2})+\sum\limits_{1}^{3} \a_{i} \hhl_{i}+\sum\limits_{i=-3, i\neq 0}^{3}\hat{\e}_i\hg^{-i}_{-1/2}+\sum\limits_{i=-3, i\neq 0}^{3}\hat{\e'}_i\hg^{-i}_{1/2}\bigg)\,,\label{gaugepmt2}
\end{eqnarray}
such that the global charge in this case is
\begin{equation}\label{gcg}
   Q(\bar{\g})= \int d \f {\G}\hat{\hat{\L}}= \int d \f \, (\a_1 \cl+\sum\limits_{i=1}^9 \r_i \cj_i+\sum\limits_{i=-3,i\neq 0}^3 \cg_i \e_i)\,.
\end{equation}
Following the same steps as in the previous case, we can derive the classical Poisson brackets. To make the spin-1 currents to have correct conformal dimension, we add Sugawara type of modification terms to the stress tensor
\begin{equation}
\cl \to \cl+\frac{\p}{2 k_{CS}}\sum\limits_{i=1}^{9} \cj_{i}\cj_{i}\,.
\end{equation}
After a shift of the zero mode and field redefinitions
\begin{eqnarray}
\nn\cl\rightarrow \cl+\frac{k_{CS}}{4\p} \,, \qquad \cg_{i} \rightarrow \bar{\cs}_{i}\,,\quad i= 1.. \,3\,, && \cg_{i} \rightarrow {\cs}_{-i}, \quad i= -1.. -3\,,
\end{eqnarray}
a mode expansion
\begin{eqnarray}
\nn \cl(\f)&=&\frac{1}{2\p}\sum\limits_{n} L_{n} e^{in\f}\,,\qquad \cj_{a}(\f)=\frac{1}{2\p}\sum\limits_{n} J^{a}_{n} e^{in\f}\,,\\
\nn \cs_{j}(\f)&=&\frac{1}{2\p}\sum\limits_{j} S^{j}_{n} e^{in\f}\,.\qquad \bar{\cs}_{j}(\f)=\frac{1}{2\p}\sum\limits_{j} \bar{S}^{j}_{n} e^{in\f}\,,
\end{eqnarray}
gives the following commutators between the generators
\begin{subequations}
\begin{eqnarray}
&&[ L_{m}, L_{n}] = (m-n) L_{m+n}+\frac{c}{12}(m^{3}-m)\delta_{m,-n}\,,  \\
&&[ L_{m}, J^{a}_{n}] = -n J^a_{m+n}\,,\\
&&[L_{m},S^{i}_{n}]=(\frac{m}{2}-n) S_{m+n}\,,\\
&&[L_{m},\bar{S}^{i}_{n}]=(\frac{m}{2}-n) \bar{S}_{m+n}\,,\\
&&[J^{a}_{m}, J^{b}_{n}] =f^{abc}J^{c}-2\, m\, k_{CS} \del^{ab}\del_{m+n}\,,\label{jj2}\\
&&[J^{a}_{m}, S^{i}_{n}] =(A^{a})^{ij} S^{j}_{m+n}\,,\\
&&[J^{a}_{m}, \bar{S}^{i}_{n}] =-(A^{a})^{ji} \bar{S}^{j}_{m+n}\,,\\
&&\{S^i_m,\bar{S}^j_n\}=2L_{m+n}\del^{i,j}+(m-n)(A^a)^{ij}J^a_{m+n}-\frac{2 \p}{k_{CS}} M_{ab}^{ij} J^aJ^b+2k_{CS} (m^2-\frac{1}{4})\del^{ij}\del_{m,-n}\,,\nn
\end{eqnarray}
\end{subequations}
where the central charge is related to $k_{CS} $ by $c=12\, \mbox{STr}(\hhl^2_0)\, (-k_{CS})=6\,k_{CS} $.
The constants $A^{a}_{ij}$ coincide with the set of the canonical generators of the $u(3)$ algebra with $f^{abc}$ being the structure constants  $[A^{a},A^{b}]=f^{abc} A^{c}$. It is not very enlightening to list values for the coefficients $M_{ab}^{ij}$, since they are coefficient of the nonlinear terms and will be modified upon solving the Jacobi identities.

Remarkably, this algebra is precisely the 2-dimensional nonlinear superconformal algebra with $u(3)$-supersymmetry \cite{bershadsky, mathieu}. We call this algebra $u(3)$-SCA. To compare with the results in \cite{mathieu}, we can use the following dictionary
\begin{eqnarray*}
A^{a} \to \l^{a}\,,\quad \cj_{a} \to W_{a}\,,\quad \cs_{i}\to G_{i}\,,\quad \bar{\cs}_{i}\to \bar{G}_{i}
\end{eqnarray*}
and the value of the parameter
\begin{equation*}
B=2\,,\qquad2k_{CS}=S=S_{0}\,.
\end{equation*}
The spin-$1$ currents form the adjoint representation of the $u(3)$ symmetry algebra, and $\cs_{i}$ and $\bar\cs_{i}$ are the fundamental and anti-fundamental representations of the $u(3)$ algebra respectively. Once again, the spin-2 field $\cl$ is a singlet of $u(3)$.

$\bullet$ {\bf The quantum algebra. }
As in the previous case, the above algebra we derived is a classical version of the algebra in \cite{bershadsky, mathieu}. We follow the same procedure to solve the Jacobi identities and since our classical solution once again coincides with the ansatz in \cite{bershadsky, mathieu}, the quantum algebra agrees with the result in \cite{bershadsky, mathieu}. The key result for us is the new quantum corrected central charge
\begin{equation}
c=\frac{6k^{2}_{CS}+13 k_{CS}+2}{k_{CS}+2}\,.
\end{equation}
For the full algebra that satisfying the Jacobi identities, we refer the reader to \cite{bershadsky, mathieu}.

\vspace{1.5mm}
{\bf Summary}: In this section, we have computed the asymptotic symmetries corresponding to the different $osp(1|2)$ embeddings. The results of this section can be summarized in the following table
\begin{table}[H]
\renewcommand{\arraystretch}{1.3}
\begin{tabular}{|c|c|c|c|}
  \hline
  embedding & asymptotic symmetry & central charge & known results \\\hline\hline
  principal embedding \eqref{principalembedding} & super-$\cw_3$ & $c=18k_{CS}$ & \cite{lprsw,ik,ito93,ahn93} \\\hline
  non-principal embedding \eqref{sl32dcp10} & $u(2|1)$-SCA & $c=6k_{CS}$ & \cite{dth} \\\hline
  non-principal embedding \eqref{sl32bcp01} & $u(3)$-SCA & $c=6k_{CS}$ & \cite{bershadsky, mathieu} \\
  \hline
\end{tabular}
\end{table}
Further notice that the two different non-principal embeddings really give two different theories. This is because
\begin{enumerate}
  \item The spectra of the asymptotic states are different. In the non-principal embedding \eqref{sl32dcp10}, which is denoted as AdS$^{(1)}$, there are states with ``wrong" statistics: the bosonic fields $\cm,~\cn$ with spin $s=1/2$ and the fermionic fields $\cx_i\,,~i=1,...,4$ with spin $s=0$. However, all fields in the non-principal embedding \eqref{sl32bcp01} have ``correct" statistics.
  \item When we were deriving the asymptotic symmetry, which describes the global symmetry of the asymptotic AdS$_3$ spacetime, we imposed different boundary conditions for the different embeddings. This is reflected in the different forms of the asymptotic AdS$_3$ connections \eqref{ggcnt0}, \eqref{ggcnt} and \eqref{ggcnt2} since their forms are determined by the boundary condition.
  \item The asymptotic symmetries of the  three different embeddings are different. This gives a hint that the dual CFTs should also be different.
\end{enumerate}

%%%%%%%%%%%%%%%%%%%
%%%%%%%%%%%%%%%%
\section{Geometry corresponding to the different embeddings}
%%%%%%%%%%%%%%%%%%%
%%%%%%%%%%%%%%%%
In the previous sections, we have studied different $osp(1|2)$ embeddings of the $sl(3|2)$ higher-spin supergravity. Physically, this means identifying different states in the higher-spin theory as the gravity sector. In other words, we have identified different fields with vielbeins and spin connections in different embeddings. Thus if we consistently turn off all fields other than the gravity sector, each embedding admits a vacuum solution of the higher-spin theory. This means the $sl(3|2)$ higher-spin supergravity possesses three vacua configurations.\fn{We will shortly show that the three vacua are different.} Each vacuum solution is obtained by solving the vacuum Einstein equation, which is the flatness condition in the Chern-Simons language
\begin{equation}\label{cseom}
d\G+\G\wedge \G=0\,, \qquad d\tilde\G+\tilde\G\wedge \tilde\G=0\,,
\end{equation}
where the gauge connection $\G$, $\tilde\G$ contains only the gravity sector with all other fields turned off. The AdS vacuum sulotion reads
\begin{eqnarray}
\G_{\text{AdS}}=e^{\r} L_{1}dx^{+}+L_{0}d\r\,,  \qquad\tilde{\G}_{\text{AdS}}=-e^{\r} L_{-1}dx^{-}-L_{0}d\r \label{adsvcm}
\end{eqnarray}
where $L_{0,\pm1}$ are the gravitational $sl(2,\br)$ generators for each $osp(1|2)$ embedding.

In this section, we will discuss the properties of the AdS vacua and the relations among them.

%%%%%%%%%%%%%%%%%%%
\subsection{AdS vacua corresponding to different embeddings}
%%%%%%%%%%%%%%%%%%%
For the three different embeddings, the generators of the $sl(2)$ sector are shown in \eqref{prinpemb}, \eqref{spin2hat}, \eqref{spin2hat2}. Plugging these into \eqref{adsvcm} and \eqref{genmetric}, the metrics of the AdS vacuum of the three corresponding embeddings all take the standard form:
\begin{equation}\label{ads1}
  ds^2=\ell_i^2 (d\r^2-e^{2\r}dx_+ dx_-)\,,\qquad i=p,1,2\,.
\end{equation}
They represent AdS$_3$ vacua with AdS radius $\ell_{AdS}$ being $|\ell_p|$, $|\ell_1|$, $|\ell_2|$ respectively.

As discussed earlier in section \ref{nmlz}, we fix the $k_{CS}$ in different embeddings, which leads to different AdS radii for different embeddings.
In our specific case, we have
\begin{equation}\label{adsradii}
  4\, G_3 \, k_{CS}=\frac{\ell_p}{\mbox{STr}(L^2_0)}=\frac{\ell_1}{\mbox{STr} (\hl^2_0)}=\frac{\ell_2}{\mbox{STr}(\hhl^2_0)}\,.
\end{equation}
Following the definition \eqref{str}, the super-traces can be computed as
\begin{equation}\label{strll}
  \mbox{STr}(L^2_0)=\frac{3}{2}\,,\qquad \mbox{STr}(\hl^2_0)=\frac{1}{2}\,,\qquad \mbox{STr}(\hhl^2_0)=-\frac{1}{2}\,.
\end{equation}
Then the AdS radii corresponding to the two non-principal embeddings are 1/3 of the AdS radius of the principal embedding.

%%%%%%%%%%%%%%%%%%%
\subsection{RG flows between different AdS vacua}
%%%%%%%%%%%%%%%%%%%
In this subsection, we will show that the theories derived from different embeddings are related by RG flows. We will identify the two non-principal embeddings as two UV theories and the principal embedding as an IR theory.
We will construct interpolation solutions from each UV vacuum to the IR vacuum. Then we verify our proposal by showing that  the solution flow to the UV vacuum at $\r\to\infty$ and to the IR vacuum at $\r\to -\infty$, which is similar to the construction in \cite{agkp}.

Before doing that, we want to address the motivation of the RG flow. We understand the holographic RG flow in the following sense
\begin{enumerate}
\item From what we will show next, we can explicitly construct solutions of the Chern-Simons equation of motion \eqref{cseom} that interpolate between different AdS vacua. This is the conventional treatment of the holographic RG flow.
\item As mentioned at the end of the previous section, different embeddings give different theories. So the interpolation solutions are not trivial field relabeling since they connect distinct theories.
\item From the dual CFT point of view, we will find operators triggering the RG flows by analyzing the asymptotics of the interpolation solutions.
\end{enumerate}
To simplify our notation, we call the vacuum corresponding to the principal embedding AdS$^{(\rm p)}$, the vacuum corresponding to the embedding \eqref{sl32dcp10} AdS$^{(1)}$ and the vacuum corresponding to the embedding \eqref{sl32bcp01} AdS$^{(2)}$.

\subsubsection{RG flow from UV vacuum AdS$^{(1)}$ to the IR vacuum AdS$^{(\rm p)}$}

To find the relation between different AdS vacua, we construct solutions of the equation of motion \eqref{cseom} interpolating between them. Unlike the bosonic case, the $sl(2)$ generators $L_0,~ \hl_0,~ \hhl_0$ of different embeddings are not proportional to each other, so the simple solution in \cite{agkp} does not hold in the current supersymmetric case.

As a result, we consider the solution
\begin{eqnarray}
\hspace{-6mm}\G& \hspace{-3mm}=& \hspace{-3mm}\m\bigg( (\frac{16 }{9}e^{e^{\r}/2+\r}-\frac{4}{9} e^{\r}) L_{1}- \frac{4}{3}(e^{e^{\r}/2+\r}-e^{\r}) K_{1}\bigg)dx^{+}+e^{e^{\r}+2\r} \hl_{1}dx^{-}+(L_{0}+ e^{\r}\hl_{0})d\r \label{1p1}\\
\nn \tilde{\G}& \hspace{-3mm}=&  \hspace{-3mm}-\m\bigg( (\frac{16 }{9}e^{e^{\r}/2+\r}-\frac{4}{9} e^{\r}) L_{-1}- \frac{4}{3}(e^{e^{\r}/2+\r}-e^{\r}) K_{-1}\bigg)dx^{-}-e^{e^{\r}+2\r} \hl_{-1}dx^{+}-(L_{0}+ e^{\r}\hl_{0})d\r\,.
\end{eqnarray}
Note that the $\m$ is an arbitrary constant parameter. In the $\r\to \infty$ limit, we have $e^{\r}\gg \r$, $e^{e^{\r}}\gg e^{e^{\r}/2}$, so the dominant terms in the above solution are
\begin{eqnarray}
\G&=& e^{e^{\r}} \hl_{1}dx^{-}+e^{\r}\hl_{0}d\r\,, \nn\\
 \tilde{\G}&=& -e^{e^{\r}} \hl_{-1}dx^{+}- e^{\r}\hl_{0}d\r\,.
\end{eqnarray}
We can define $\tilde{\r}=e^{\r}$, then the above result reads
\begin{eqnarray}
\G&=& e^{\tilde{\r}} \hl_{1}dx^{-}+\hl_{0}d\tilde{\r}\,, \nn\\
 \tilde{\G}&=& -e^{\tilde{\r}} \hl_{-1}dx^{+}-\hl_{0}d\tilde{\r}\,,\label{1plarger}
\end{eqnarray}
which is the standard AdS vacuum \eqref{adsvcm} after switching $x^{+}$ with $x^{-}$. This means the interpolation solution \eqref{1p1} approaches to the AdS$^{(1)}$ vacuum in the UV.

In the $\r\to -\infty$ limit, we have $e^{\r}\to 0$, $e^{\r}\gg e^{2\r}$, so the solution \eqref{1p1} reads
\begin{eqnarray}
\G&=& \frac{4}{3}\,\m\, e^{\r} L_{1}dx^{+}+L_{0}d\r\,, \nn\\
 \tilde{\G}&=& - \frac{4}{3}\m\, e^{\r} L_{-1}dx^{-}-L_{0}d\r\,.\label{1psmallr}
\end{eqnarray}
Then a shift $\r\to \tilde\r=\r+\ln\frac{4\,\m}{3}$ will take it to the standard form \eqref{adsvcm}. This means the interpolation solution \eqref{1p1} approaches the AdS$^{(\rm p)}$ vacuum in the IR.

Further, notice that the interpolation only flows in one direction: we cannot find a solution that approaches the AdS$^{(\rm p)}$  at large $\r$ and approaches to AdS$^{(1)}$ vacuum at small $\r$.\fn{For example, if we naively change $e^{\r}$ to $e^{-\r}$ in the $d\r$ term of \eqref{1p1}, then the solution does not flow to the desired AdS vacuum solution at least in one of the two limits.}

\subsubsection{RG flow from UV vacuum AdS$^{(2)}$ to the IR vacuum AdS$^{(\rm p)}$}\label{rg2p}

For this case, we consider the solution\fn{We do not find interpolation solutions with UV and IR generators in different components, namely with both $dx^{+}$ and $dx^{-}$ non-vanishing as \eqref{1p1}.}
\begin{eqnarray}
\G&=& \bigg( \frac{1}{3}(4 e^{\m\r}- e^{e^{\r}+\m\r}) L_{1}- (e^{\m\r}-e^{e^{\r}+\m\r}) K_{1}\bigg)dx^{+}+(\m L_{0}+e^{\r}\hhl_{0})d\r\,, \nn\\
 \tilde{\G}&=&  -\bigg(\frac{1}{3}(4 e^{\m\r}- e^{e^{\r}+\m \r}) L_{-1}- (e^{\m\r}-e^{e^{\r}+\m\r}) K_{-1}\bigg)dx^{-}-(\m L_{0}+e^{\r}\hhl_{0})d\r\,.\label{2p}
\end{eqnarray}
In the limit $\r\to \infty$, the above solution approaches to
\begin{eqnarray}
\G&=&  \bigg( -\frac{1}{3} e^{e^{\r}} L_{1}+ e^{e^{\r}} K_{1}\bigg)dx^{+}+e^{\r}\hhl_{0}d\r\,, \nn\\
 \tilde{\G}&=&   \bigg( \frac{1}{3} e^{e^{\r}} L_{-1}- e^{e^{\r}} K_{-1}\bigg)dx^{-}-e^{\r}\hhl_{0}d\r \,,\end{eqnarray}
Notice that since $\hhl_{\pm 1}=K_{\pm1}-\frac{1}{3}L_{\pm 1}$, we can rewrite the above result as
\begin{eqnarray}
\G&=&   e^{e^{\r}} \hhl_{1}dx^{+}+e^{\r}\hhl_{0}d\r\,, \nn\\
 \tilde{\G}&=&  -e^{e^{\r}} \hhl_{-1}dx^{-}-e^{\r}\hhl_{0}d\r \,,\label{2plarge}
\end{eqnarray}
which gives the standard form \eqref{adsvcm} after the redefinition  $\tilde{\r}=e^{\r}$.

In the limit $\r \to -\infty$, the solution \eqref{2p} approaches to
\begin{eqnarray}
\G&=& e^{\m\r} L_{1}dx^{+}+\m\,L_{0}d\r\,, \nn\\
 \tilde{\G}&=&  -  e^{\m\,\r} L_{-1}dx^{-}-\m\,L_{0}d\r\,.
\end{eqnarray}
Then a rescaling $\r\to \tilde{\r}=\m\r$ gives the standard gauge connection for the AdS vacuum
\begin{eqnarray}
\G&=& e^{\tilde\r} L_{1}dx^{+}+\,L_{0}d\tilde\r \nn\\
 \tilde{\G}&=&  -  e^{\tilde\r} L_{-1}dx^{-}-\,L_{0}d\tilde\r\,.\label{2psmall}
\end{eqnarray}

As shown in the previous subsection, a key point for us to understand this solution as an RG flow is that it is a one-direction flow: we do not find a solution which approaches the AdS$^{(\rm p)}$ vacuum in the $\r \to \infty$ limit and approaches the AdS$^{(2)}$ vacuum in the $\r \to -\infty$ limit.

\subsubsection{Interpolations between the two UV vacua AdS$^{(1)}$ and AdS$^{(2)}$}\label{uvuv}
We can also find interpolation solutions between AdS$^{(1)}$ and AdS$^{(2)}$. But they should not be understood as RG flows: there are solutions interpolating in both the two directions. Consider the following interpolation solution
\begin{eqnarray}
\nn \G&=&e^{a(\r)}\hl_1 dx^+  +e^{b(\r)}\hhl_1 dx^-+\big(a'(\r) \hl_0+b'(\r) \hhl_0\big)d\r\,,\\
 \tilde{\G} &=&-e^{a(\r)}\hl_{-1} dx^+  -e^{b(\r)}\hhl_{-1} dx^- -\big(a'(\r) \hl_0+b'(\r) \hhl_0\big)d\r\,,\label{int23}
\end{eqnarray}
where $a(\r)$ and $b(\r)$ are arbitrary functions of $\r$.\\
$\bullet$ ~~Interpolating from AdS$^{(1)}$ to AdS$^{(2)}$ as $\r$ decreases.

Now consider $a(\r)=e^{\r} $ and $b(\r)=e^{-\r} $, then at $\r \to \infty$, ${a(\r)} \gg {b(\r)}$, $e^{a(\r)} \gg e^{b(\r)}$ and the solution \eqref{int23} approaches AdS$^{(1)}$ after a field redefinition $\r\to\tilde\r=e^{\r}$
\begin{eqnarray}
\nn \G&=&e^{\tilde\r}\hl_1 dx^+ + \hl_0 d \tilde\r\,,\\
 \tilde{\G} &=&-e^{\tilde\r}\hl_{-1} dx^+  - \hl_0d\tilde\r \,. \label{int23large1}
\end{eqnarray}

In the $\r \to -\infty$ limit, $a(\r) \ll b(\r)$, $e^{a(\r)} \ll e^{b(\r)}$ and the solution \eqref{int23} approaches AdS$^{(2)}$ after a field redefinition $\r\to\tilde\r'=e^{-\r}$
\begin{eqnarray}
\nn \G&=&e^{\tilde\r'}\hl_1 dx^+ + \hl_0 d \tilde\r'\,,\\
 \tilde{\G} &=&-e^{\tilde\r'}\hl_{-1} dx^+  - \hl_0d\tilde\r' \,. \label{int23small1}
\end{eqnarray}
$\bullet$ ~~Interpolating from AdS$^{(2)}$ to AdS$^{(1)}$ as $\r$ decreases.

Now consider $a(\r)=e^{-\r} $ and $b(\r)=e^{\r} $, then at $\r \to \infty$, $a(\r) \ll b(\r)$, $e^{a(\r)} \ll e^{b(\r)}$ and the solution \eqref{int23} approaches AdS$^{(2)}$ after a field redefinition $\r\to\tilde\r=e^{\r}$
\begin{eqnarray}
\nn \G&=&e^{\tilde\r}\hl_1 dx^+ + \hl_0 d \tilde\r\,,\\
 \tilde{\G} &=&-e^{\tilde\r}\hl_{-1} dx^+  - \hl_0d\tilde\r \,. \label{int23large2}
\end{eqnarray}

In the limit $\r \to -\infty$, $a(\r) \gg b(\r)$, $e^{a(\r)} \gg e^{b(\r)}$ and the solution \eqref{int23} approaches AdS$^{(1)}$ after a field redefinition $\r\to\tilde\r'=e^{-\r}$
\begin{eqnarray}
\nn \G&=&e^{\tilde\r'}\hl_1 dx^+ + \hl_0 d \tilde\r'\,,\\
 \tilde{\G} &=&-e^{\tilde\r'}\hl_{-1} dx^+  -\hl_0d\tilde\r'  \,. \label{int23small1}
\end{eqnarray}

The origin of the existence of interpolations in both the two directions is the following commutators
\begin{equation}\label{keycmt}
  [\hl_0,\hhl_{\pm 1}]=0\,, \qquad [\hhl_0,\hl_{\pm 1}]=0\,.
\end{equation}
This means the $sl(2,\br)$ generators of these two embeddings decouple and the solution \eqref{int23} is really a ``direct sum"
of two mutually commuting solutions
\begin{equation}\label{int23re}
  \G=\G^{(1)}+\G^{(2)}\,,\qquad \tilde{\G}=\tilde{\G}^{(1)}+\tilde{\G}^{(2)}\,,
\end{equation}
where
\begin{eqnarray*}
 && \G^{(1)}=e^{a(\r)}\hl_1 dx^+ +a'(\r) \hl_0 d\r\,,\quad \G^{(2)}= e^{b(\r)}\hhl_1 dx^-+b'(\r) \hhl_0 d\r\,,\\
&&\tilde{\G}^{(1)}=-(e^{a(\r)}\hl_{-1} dx^+ +a'(\r) \hl_0 d\r)\,,\quad \tilde{\G}^{(2)}= -(e^{b(\r)}\hhl_{-1} dx^-+b'(\r) \hhl_0 d\r)\,.
\end{eqnarray*}
In this language, the commutators \eqref{keycmt} translate to
\begin{equation}\label{muturalcnt}
  \G^{(1)}\wedge \G^{(2)}=0\,,
\end{equation}
which allows $\G^{(1)},\,\G^{(2)}$ and $\G$ to be flat connections simultaneously. This gives another piece of evidence showing that the two non-principal embeddings give two different theories at UV.\fn{This is because the two UV theories being the same means $\G^{(1)}\sim \G^{(2)}$ up to relabeling, then the flat condition (or equation of motion) should read $d\G^{(1)}+\G^{(1)}\wedge\G^{(2)}=0$ with an extra $d\G^{(1)}$ term.} We are free to tune $a(\r)$ and $b(\r)$ independently, and as shown above, we can interpolate between the two AdS vacua in both directions with special choices of $a(\r)$ and $b(\r)$.

Actually, we can further rewrite \eqref{int23re} in the following way
\begin{eqnarray}\label{int23resum}
 && \G^{(1)}=e^{a(\r)}\hl_1 dx^+ +\hl_0 d\, a(\r)\,,\quad \G^{(2)}= e^{b(\r)}\hhl_1 dx^-+ \hhl_0 d\, b(\r)\,,\\
&&\tilde{\G}^{(1)}=-(e^{a(\r)}\hl_{-1} dx^+ + \hl_0 d\, a(\r))\,,\quad \tilde{\G}^{(2)}= -(e^{b(\r)}\hhl_{-1} dx^-+ \hhl_0 d\, b(\r))\,.
\end{eqnarray}
Hence we see that $(A^{(1)},\bar A^{(1)})$ describes an AdS$_3$ vacuum with radial coordinate $\tilde\r=a(\r)$, and $(A^{(2)},\bar A^{(2)})$ describes an AdS$_3$ vacuum with radial coordinate $\tilde\r=b(\r)$.
Therefore the solution \eqref{int23} is a trivial combination of two independent solutions and do not behave as a standard RG flow.

Alternatively, the existence of interpolations in both the two directions might be understood as an RG cycle. As we will show in section \ref{opttr}, the RG flows in our discussion are triggered by Lorentz-violating operators so the c-theorem, which requires Lorentz invariance, does not forbid the existence of such an RG cycle.\fn{We thank Thomas Hartman for suggesting this interpretation to us.} Note that  our observation in this section does not behave in the same way as the limit cycles in 4D \cite{fgs}: our interpolation is constructed from two RG flows, each ending at a CFT. In addition, we have to turn on chemical potentials to trigger each of the constituent RG flows. Therefore,  the conclusion in \cite{lpr} that the limit cycles are equivalent to conformal fixed points  does not apply to our observation.

{\bf Summary}: From the explicit computations we have done in this section, we show that the two non-principal embeddings give two UV theories that flow to the IR theory corresponding to the principal embedding. Thus we find a duality between the two UV theories in the sense that they flow to the same IR theory provided that some chemical potentials or operator perturbations are turned on.

\subsection{Comparison with the bosonic spin-$3$ gravity}
In this subsection, we want to investigate the relation between our results and the results in \cite{agkp}, which we have summarized in section \ref{rvbos}. It is well understood that the principal embedding in our discussion is the $\cn=2$ supersymmetric extension of the principal embedding into the $sl(3)$ algebra. We claim that our non-principal embedding I, namely the AdS$^{(1)}$, in section \ref{non1} is a direct supersymmetrization of the diagonal embedding in \cite{agkp}. The evidence is as follows
\begin{enumerate}
  \item From the embedding itself, we can truncate our non-principal embedding I in \ref{non1} to the diagonal embedding of $sl(3)$ discused in \cite{agkp}. We can truncate our result in two steps: first, turn off all the fermionic part of the $sl(3|2)$ superalgebra, which means we neglect \eqref{sl32fdcp10}; second, truncate out the $sl(2)\oplus u(1)$ part of the bosonic subalgebra \eqref{bs32}, which means we drop the $3\cd'_0\oplus \cd_0$ representations of the decomposition \eqref{sl32bdcp10}. Then the resulting decomposition of the $sl(3)$ algebra from the truncation of \eqref{sl32bdcp10} reads $sl(3)=\cd_1\oplus 2\tilde{\cd}_{1/2}\oplus \cd_0$. This is precisely the decomposition obtained from the diagonal embedding of $sl(3)$ discussed in \cite{agkp}.
  \item The RG flow solution \eqref{1p1} is similar to the RG solution in the bosonic case in the sense that the UV generators and the IR generators sit in opposite-chirality-components of the solution. In other words, the chemical potential term is separable from the AdS$_3$ vacuum solution in the UV.
\end{enumerate}

While using a similar reasoning, we conclude that our non-principal embedding II, namely AdS$^{(2)}$, in section \ref{non2} is a new result due to the presence of supersymmetry and has no counterpart in the bosonic case. The argument goes as
\begin{enumerate}
  \item If we consider a similar truncation of the non-principal embedding II \eqref{sl32bcp01}, we reach the decomposition $sl(3)=8\cd_0$. This is not observed in the bosonic case since this embedding is trivial and no spin-$1$ representation of the $sl(2)$, who has spacetime spin-$2$, is present. The resulting theory will not contain the gravity sector and thus should not be considered in the discussion in \cite{agkp}.
  \item The RG solution \eqref{2p} looks very different from that in the bosonic case since there is only a $dx^+$ component and the perturbations are not manifestly separable as in the bosonic case.
\end{enumerate}

Therefore, we conclude that the non-principal $osp(1|2)$ embedding of $sl(3|2)$ in section \ref{non1}, i.e.~AdS$^{(1)}$, is a simple supersymmetric extension of the diagonal $sl(2)$ embedding of $sl(3)$ in \cite{agkp}. While the other non-principal $osp(1|2)$ embedding of $sl(3|2)$ in section \ref{non2}, i.e.~AdS$^{(2)}$, is brought to us purely by supersymmetry. As a result, the duality we have claimed at the end of the previous subsection is also a bonus relation of supersymmetry.

The origin of this duality is not clear at this moment, but it must have something to do with the structure of the Lie superalgebra $sl(n|n-1)$. The bosonic subalgebra of $sl(n|n-1)$ lookes like $sl(n|n-1)\big|_B=sl(n)\oplus sl(n-1)\oplus u(1)$. And the two sides of the duality correspond to identifying the (bosonic) gravitational $sl(2)$ subalgebra from $sl(n)$ and $sl(n-1)$, respectively. This is at least true for the $n=3$ case. We would like to study this structure in full detail in the future.
\subsection{Operators generating the RG flows}\label{opttr}

The above interpolation solutions can be obtained by adding terms to the AdS vacua solutions at UV and solving the equation of motion. Those extra terms trigger the RG flows.

\vspace{1mm}
\noindent$\bullet$ {\bf  The operators triggering the RG flow from AdS$^{(1)}$ to  AdS$^{(\rm p)}$.}

From UV point of view,  the RG flow is triggered by the $\m$ term in \eqref{1p1}. In the bases of the UV theory \eqref{othebd1b}, the $\m$ term reads
  $\m(\frac{8 \sqrt{2}}{3}e^{e^\r/2+\r}(\hm_{1/2}+\hn_{1/2})-\frac{4 }{3}e^\r \,i\hj_3)$,
which is real and the ``$i$" shows up because ``$\hj_3$" is defined to be imaginary. However, this is not the final result due to the explicit $\r$ dependence. The correct expression for the UV theory should be obtained by taking the $\r \to \infty$ limit, where the leading terms are
 $ \m\,\frac{8 \sqrt{2}}{3}e^{e^\r/2}(\hm_{1/2}+\hn_{1/2})$,
with spin-$3/2$ generators
$\hm_{1/2}+\hn_{1/2}$\,.
So the RG flow is triggered by adding this term to the AdS vacuum solution, which corresponds to adding ``bosonic'' spin-$3/2$ currents to the Lagrangian. This is very similar to the observation in the bosonic case \cite{agkp}.

But the situation is a little more involved in the current supersymmetric case. First of all, in the bosonic case  \cite{agkp} the $\hl_0$ generator in the UV theory is proportional to the $L_0$ generator in the IR theory. So the interpolation solution (2.27) in \cite{agkp} has a simple $d\r$ term. But in the supersymmetric case, the $\hl_0$ generator in the UV theory is not proportional to the $L_0$, therefore from the solution \eqref{1p1}, we have to include $L_0$ in $d\r$ terms, which corresponds to $\frac{1}{2}i(\hj_1-\hj_4)+2\hl_0$ in the UV bases \eqref{othebd1b}. Thus in order to construct the interpolation solution, we have to turn on some spin-$1$ fields corresponding to $\hj_1-\hj_4$ as well.  Secondly, in the $\m$ dependent term, there is another spin-1 field $\hj_{3}$ turned on.  Although it is suppressed in the UV limit $\r \to \infty$, it is still required to solve the equation of motion.

In summary, in addition to the spin-$3/2$ fields,  the spin-1 currents corresponding to $\hj_{1},\, \hj_{3},\, \hj_{4}$ are also needed to solve the equation of motion.

\vspace{1mm}
\noindent$\bullet$ {\bf The operators triggering the RG flow from AdS$^{(2)}$ to  AdS$^{(\rm p)}$.}

As discussed in section \eqref{rg2p}, there is no solution that separate the chemical potential, i.e.~$\m$, term from  the AdS vacuum term. So the fields triggering the flow do not stand separately as in the previous case. Inspired from the analysis of the previous RG flow, we can capture the triggering fields by investigating the sub-leading fields in the UV limit $\r \to \infty $. From the form of the solution \eqref{2p}, we can see that in the $d\r$ term we need to turn on field $L_{0}$, which reads   $\hhl_{0}+\hhj_{3}+\sqrt{3}\hhj_{8}$ in the second UV bases \eqref{othebd2}. Although it is suppressed at $\r \to \infty $, this term will become more and more important as we flow to the IR theory and will be dominant there. Besides, in the UV limit $\r \to \infty $ the $dx^{+}$ term also possesses a subleading term $\frac{4}{3}\, e^{\m\r} L_{1}-e^{\m\r} K_{1}$, which reads $-2(\hhj_{1}-i\,\hhj_{2}-\hhj_{6}+i\,\hhj_{7})e^{\m\r}$ in the second UV bases \eqref{othebd2}. Again due to the explicit $\r$ dependence, this subleading term in the UV becomes more and more relevant when flowing to the IR theory.

 In summary, to trigger the flow, we need to  add  subleading terms proportional to $\hhj_{3}+\sqrt{3}\hhj_{8},\, \hhj_{1}-i\,\hhj_{2}-\hhj_{6}+i\,\hhj_{7}$ in the UV AdS vacuum. This corresponds to turning on some certain combinations of the spin-$1$ currents in the Lagrangian. From the UV point of view, the $\m$ parameter sources the $L_{0}$ generator in the $d\r$ term of the interpolation solution \eqref{2p} as a chemical potential. However,  it enters the $dx^{+}$ term of the solution in an unfamiliar exponential way. We  hope to investigate the role of $\m$ in detail in the future. Nevertheless, as we will see in the next section, the $\m$ drops out when we try to relate fields in the UV theory AdS$^{(2)}$ and the fields in the IR theory AdS$^{(\rm p)}$.

\subsection{Relations between UV and IR operators}

As discussed in the previous subsection, we can construct interpolation solutions between UV theories and the IR theory.  In this subsection, we want to find what fields in the IR theory do UV fields flow to. In practice,  we consider the perturbations around the interpolation solutions. By solving the linearized equation of motion,\fn{The equation of motion of the perturbed RG flow from AdS$^{(2)}$ to AdS$^{(\rm p)}$ is linear due to the simple form of \eqref{2p}, therefore linearization is not necessary.} we get the relations between UV and IR fields.

In the case we are considering, the $L_0,\,\hl_{0}\,\hhl_{0}$ generators in different embeddings are not proportional to each other. So the interpolation solutions have complicated $d\r$ terms. A consequence of this complication is that the $\r$ dependence of various fields are not separable in general:  we cannot factor out the $\r$ dependence for each field as in the bosonic case \cite{agkp}. However, as we will see below, this non-trivial $\r$ dependence allows us to keep track of how do different fields mix along the flow, which is not observed in the bosonic case.

\subsubsection{Relations between operators in AdS$^{(1)}$ and AdS$^{(\rm p)}$}\label{optsec1p}

We start by perturbing the RG flow, i.e.~the interpolation solution \eqref{1p1}, by adding all possible fields in the highest weight gauge
\begin{eqnarray}
  \hspace{-6mm}\G& \hspace{-3mm}=& \hspace{-3mm}\m\bigg( (\frac{16 }{9}e^{e^{\r}/2+\r}-\frac{4}{9} e^{\r}) L_{1}- \frac{4}{3}(e^{e^{\r}/2+\r}-e^{\r}) K_{1}\bigg)dx^{+}+e^{e^{\r}+2\r} \hl_{1}dx^{-} \\
  &&+ (L_{0}+ e^{\r}\hl_{0})d\r+(\cl'_{IR} L_{-1}+\ca'_{IR} K_{-1}+\cw'_{IR} W_{-2}+\cj'_{IR} J ) dx^+\\
  &&+ (\cl'_{UV} \hl_{-1}+\cp'_{UV} \hm_{-1}+\cq'_{UV} \hn_{-1}+\sum\limits_{i=1}^5 \cj'_{UV} \hj^i_{-1}) dx^-
\end{eqnarray}
where all the fields $\co'_{IR}\, (\co'_{UV})$ depend on $(x^+,\,x^{-},\,\r)$. The $\tilde{\G}$ connection is similar so we only focus on the $\G$ connection.

Solving the $dx^+d\r$ and $dx^-d\r$ components of the equation of motion respectively, we work out the explicit $\r$ dependence of various fields
\begin{eqnarray}
\nn\hspace{-6mm}\G^{(1p)}& \hspace{-3mm}=& \hspace{-3mm}\m \bigg( (\frac{16 }{9}e^{e^{\r}/2+\r}-\frac{4}{9} e^{\r}) L_{1}- \frac{4}{3}(e^{e^{\r}/2+\r}-e^{\r}) K_{1}\bigg)dx^{+}+e^{e^{\r}+2\r} \hl_{1}dx^{-} \\
  \nn&&+ (L_{0}+ e^{\r}\hl_{0})d\r+\bigg(\cj_{IR} J+\frac{1}{9} \big(4 e^{-\frac{e^\r}{2}-\r} (\ca_{IR}+3 \cl_{IR})-e^{-\r} (4 \ca_{IR}+3 \cl_{IR})\big) L_{-1}\\
 \nn &&+\frac{1}{3} \big(e^{-\r} (4 \ca_{IR}+3 \cl_{IR})-4 e^{-\frac{e^\r}{2}-\r} (\ca_{IR}+3 \cl_{IR})\big) K_{-1}+\cw_{IR}e^{-e^\r-2 \r} W_{-2} \bigg) dx^+\\
 \nn &&+ \bigg(e^{-e^\r-2 \r}\hat\cl_{UV} \hl_{-1}+e^{-\frac{e^\r}{2}-\r}\cm_{UV} \hm_{-1}+e^{-\frac{e^\r}{2}-\r}\cn_{UV} \hn_{-1}\\
  &&+\sum\limits_{i=1,4,5} \hat\cj^{i}_{UV} \hj^i_{-1}+e^{-\r}\sum\limits_{i=2,3} \hat\cj^{i}_{UV} \hj^i_{-1}\bigg) dx^-\,.\label{1ppert}
\end{eqnarray}
Now all the fields $\co_{IR}\, (\co_{UV})$ depend on $(x^+,\,x^{-})$ but not on $\r$. Furthermore, in the UV limit $\r\to \infty$, the $dx^{-}$ term in \eqref{1ppert} reduces to the bosonic part of the asymptotic AdS$^{(1)}$ vacuum \eqref{ggcnt} up to normalization factor $\frac{2 \p}{k_{CS}}$. In the IR limit $\r\to -\infty$, the $dx^{+}$ term in \eqref{1ppert} reduces to the bosonic part of the asymptotic AdS$^{(\rm p)}$ vacuum \eqref{ggcnt0} up to normalization factor $\frac{2 \p}{k_{CS}}$.

Now consider the linearized $dx^+ dx^-$ component of the equation of motion, which gives the following relations between the IR operators and the UV operators
\begin{subequations}\label{1poptmap}
\begin{eqnarray}
\hat\cl_{UV}&=&-3\,\frac{\pa_{+}^{2}(\ca_{IR}+3\, \cl_{IR})}{512 \m^{3}}\,,\\
\cm_{UV}&=&\sqrt{2}\,\frac{\pa_{+}(\ca_{IR}+3\, \cl_{IR}) +24 \m\, \cw_{IR}}{64 \m^{2}}\,,\label{irw2m}\\
\cn_{UV}&=&\sqrt{2}\,\frac{\pa_{+}(\ca_{IR}+3\, \cl_{IR})-24 \m\, \cw_{IR}}{64 \m^{2}}\,,\label{irw2n}\\
\hat\cj^{1}_{UV}+\hat\cj^{4}_{UV}+3\,\hat\cj^{5}_{UV}&=&\frac{-i}{\m}(\ca_{IR}+3\, \cl_{IR})\,,\label{irl2}\\
\pa_{+}(\hat\cj^{1}_{UV}+\hat\cj^{4}_{UV})&=&6i\,\pa_{-}\cj_{IR}\,,\\
\pa_{+}\hat\cj^{2}_{UV}&=&i\,\big(\frac{3}{2\m}e^{-e^{\r}/2}\pa_{+}\cw_{IR}+(1-4e^{-e^{\r}/2})\pa_{-}\cl_{IR}+\frac{4}{3}(1-e^{-e^{\r}/2})\pa_{-}\ca_{IR} \big)\,,\nn\\\label{j2run}
\end{eqnarray}
\end{subequations}
as well as the relations
\begin{subequations}\label{1pwdid}
\begin{eqnarray}
\pa_{+} \hat\cl_{UV}&=&\frac{2\sqrt{2}\m}{3}\,\pa_{-}(\cm_{UV}-\cn_{UV })\,,\\
\pa_{+}(\cm_{UV}-\cn_{UV })&=& \frac{2\sqrt{2}\m}{3i}\, \pa_{-} (\hat\cj^{1}_{UV}+\hat\cj^{4}_{UV}+3\,\hat\cj^{5}_{UV})\,,\\
 \pa_{+}\hat\cj_{UV}^{3}&=&\frac{4\m}{3}(\hat\cj^{1}_{UV}-\hat\cj^{4}_{UV})\,,\\
 \pa_{+}(\hat\cj^{1}_{UV}-\hat\cj^{4}_{UV})&=&-\frac{8\m}{3} \hat\cj_{UV}^{2}\,.
\end{eqnarray}
\end{subequations}
The various factors ``$i$'' in the above expressions are not essential, they come from our complexification of the generators \eqref{othebd1b}, we can absorb all the ``$i$'' in \eqref{othebd1b} and the results will no longer involve any ``$i$''. In addition, we see a mixing between $\ca_{IR}$ and $\cl_{IR}$ in both \eqref{1ppert} and \eqref{1poptmap}. This is a result of the $\cn=2$ supersymmetry that produces another set of spin-$1$ generators $K_{i}$ besides the $L_{i}$. The fact that the commutation relations $[L_{I}, \cdot]$ and  $[K_{I}, \cdot]$, where $\cdot$ stands for other generators in the algebra, are almost identical directly leads to the mixing observed above.

From \eqref{irl2}, we see that the IR field $\cl_{IR}$ is locally related to spacetime spin-1 fields in the UV, which agrees with the observation in the bosonic case \cite{agkp}.  From \eqref{irw2m} and  \eqref{irw2n}, we see the IR spin-3 field is related to a certain combination of the spin-$3/2$ fields in the UV by $\cw_{IR}=\frac{2\sqrt{2}\m}{3}\,(\cm_{UV}-\cn_{UV })$, which is a reasonable extension of the observation in the bosonic case \cite{agkp}.

Equation \eqref{1poptmap} also tells us how do the dimensions of various UV operators change along the flow. The operators $\hat\cl_{UV},\, \cm_{UV},\, \cn_{UV},\, \cj_{UV}^{i}$ have conformal dimension $4,3,3,2$ in the IR respectively. The conformal dimensions of these fields are all doubled, we do not know if this is accidental or not.

Equation \eqref{1pwdid} represents the chiral conservation of the UV currents $\cl_{UV},\, (\cm_{UV}-\cn_{UV }),\, \hat\cj_{UV}^{3},\, (\hat\cj^{1}_{UV}-\hat\cj^{4}_{UV})$. This is because $\m$ is turned off when we perturb around the UV vacua (i.e.~in the UV limit), which makes the right hand side of equation \eqref{1pwdid} vanish identically.  From the conservation laws, our choice of the generators \eqref{othebd1b} does not seem to be a best one when studying the behaviors of different operators. A certain recombination, namely $\check\cj_{UV}=\hat\cj^{1}_{UV}-\hat\cj^{4}_{UV},~ \check\cm_{UV}=(\cm_{UV}-\cn_{UV })$ might be more fundamental since their conservations are explicit in this analysis.

In addition to the agreement with the bosonic case, the unconventional $\r$ dependence is not observed in the bosonic case. This is because the relation $\hl_0=L_0/2$ there leads to very simple $\r$ dependences. Although our result seems a little complicated, it reveals the explicit $\r$ dependences of various fields and hence tells explicitly how fields evolve along the flow. A good example is \eqref{j2run}, where $\r$ is an explicit parameter characterizing the running of the UV (spacetime) spin-1 current $\hat\cj_{UV}^{2}$. The relations between the UV and the IR operators shown in \eqref{1poptmap} as well as those in the bosonic case \cite{agkp} can be regarded as expansions of the UV operators in the set of bases of the IR operators. The special feature of \eqref{j2run} is that the expansion coefficients change along the flow, representing the renormalization of the current. Concretely, in the UV limit $\r\to \infty$, we have
\begin{equation}
\pa_{+}\hat\cj^{2}_{UV}=i\,\pa_{-}(\cl_{IR}+\frac{4}{3}\ca_{IR} )\,.
\end{equation}
But this is not really the relation between UV operator $\pa_{+}\hat\cj^{2}_{UV}$ and the IR operators since the relation  is established at the UV scale and the $\cl_{IR},\ca_{IR}$ are merely some basis vectors but not the real IR operators. We have to run down to the IR scale $\r \to -\infty$\fn{In addition, we take $\m\to \infty$ in the IR. This can be seen from the form of the IR limit \eqref{1psmallr} that $\m$ has to be large in order for the $dx^{\pm}$ term to be non-vanishing.}, then the relation becomes
\begin{equation}
\pa_{+}\hat\cj^{2}_{UV}=-3i\pa_{-}\cl_{IR}\,,\label{j2runir}
\end{equation}
 which can be interpreted as the true relation between the UV and IR operators. Note that we do not see similar runnings in either the bosonic case \cite{agkp} or the other relations in \eqref{1poptmap} because of the canonical $\r$ dependence. This $\r$ dependent way to match the UV and IR operators will be used again in the next section.

\subsubsection{Relations between operators in AdS$^{(2)}$ and AdS$^{(\rm p)}$}\label{optsec2p}

We still want to perturb the RG flow, i.e.~the interpolation solution \eqref{2p}, by adding all possible fields in the ``highest weight'' gauge. However, unlike the previous case, the UV part and the IR part do not sit in different components ($dx^{+}$ and $dx^{-}$) in \eqref{2p}. So the simple perturbation method we used in the previous section does not work here.

To deal with this, we only add perturbation terms in the UV vacuum to the interpolation solution \eqref{2p}. Solving the equation of motion gives a general $\r$ dependent perturbed interpolation function between the UV and IR. We then run this interpolation function down to the IR, and compare this running result with the perturbed asymptotic AdS$^{(\rm p)}$ gauge connection in the ``highest weight'' gauge. This matching gives relations between the fields in the UV and IR. We will demonstrate how this works now.

We perturb around  \eqref{2p} as
\begin{eqnarray}
\G^{(2p)}&=& \bigg( \frac{1}{3}(4 e^{\m\r}- e^{e^{\r}+\m\r}) L_{1}- (e^{\m\r}-e^{e^{\r}+\m\r}) K_{1}\bigg)dx^{+}+(\m L_{0}+e^{\r}\hhl_{0})d\r \nn\\
&&+  (\tilde{\cl}'_{UV} \hhl_{-1}+\sum\limits_{i=1}^9 \tilde\cj_{UV}^{'i}\, \hhj_i)dx^+\,,\label{uvpert}
\end{eqnarray}
where all the fields $\co'_{IR}\, (\co'_{UV})$ depend on $(x^+,\,x^{-},\,\r)$. Solving the equation of motion gives the following perturbed interpolation solution
\begin{eqnarray}
\G^{(2p)}&=& \bigg( \frac{1}{3}(4 e^{\m\r}- e^{e^{\r}+\m\r}) L_{1}- (e^{\m\r}-e^{e^{\r}+\m\r}) K_{1}\bigg)dx^{+}+(\m L_{0}+e^{\r}\hhl_{0})d\r \nn\\
&&+
((C_1 e^{-\m\r}+\frac{i}{2}C_2 e^{\m\r}-(e^{-e^\r-\m \r}\tilde{\cl}_{UV}+\tilde\cj_{UV}^{6} e^{-\m\r}))\hhj_1\nn\\
&&+i(C_1 e^{-\m\r}-\frac{i}{2} C_2 e^{\m\r}-(e^{-e^\r-\m \r}\tilde{\cl}_{UV}+\tilde\cj_{UV}^{6} e^{-\m\r}))\hhj_2\nn\\
&&+(C_3-\sqrt{3}C_{8})\hhj_3
+(\tilde\cj_{UV}^{4} \cosh(2\m\r)+i C_5 \sinh(2\m\r))\hhj_4+\tilde{\cl}_{UV}e^{-e^\r+\m\r} \hhl_{-1}\nn\\
&&+(C_5 \cosh(2\m\r)-i \tilde\cj_{UV}^{4} \sinh(2\m\r))\hhj_5
+(e^{-e^\r-\m\r}\tilde{\cl}_{UV}+\tilde\cj_{UV}^{6} e^{-\m\r}+\frac{i}{2}C_7\,e^{\m\r})\hhj_6\nn\\
&&+i(e^{-e^\r-\m\r}\tilde{\cl}_{UV}+\tilde\cj_{UV}^{6} e^{-\m\r}-\frac{i}{2}C_7\,e^{\m\r})\hhj_7
+C_{8}\, \hhj_8+\tilde\cj_{UV}^{9}\, \hhj_9)\,dx^+ \,,\label{2pintuv}
\end{eqnarray}
where the $C_1,C_2,C_3,C_5,C_7,C_{8}$ are integration constant to be determined later.  Now we run to the IR theory along the perturbed flow, which is achieved by taking the $\r\to -\infty$ limit.\fn{This effectively sets $e^{\r}=0$. But we keep all the $e^{\m\r}$  untouched since the unspecific constant $\m$ can be either positive or negative.} To compare with the perturbed AdS vacuum \eqref{ggcnt0} in the IR, we further rewrite this expression in the set of bases \eqref{prinpemb} in the IR theory, which gives
\begin{eqnarray}
\G^{(2p)}&=& e^{\m \r} L_{1}dx^{+}+\m L_{0}d\r \nn\\
&&+
\bigg((e^{-\m\r}\tilde{\cl}_{UV}+\tilde\cj_{UV}^{6} e^{-\m\r}-\frac{1}{2}C_1\,e^{-\m\r})K_{-1}-\frac{C_{3}}{4}K_{0}-\frac{i}{8}e^{\r\m}(C_{7}-C_{2})K_{1}\nn\\
&&+
(-\frac{1}{3}e^{-\m \r}\tilde{\cl}_{UV}-\frac{4}{3}\tilde\cj_{UV}^{6} e^{-\m\r}+\frac{2}{3}C_1\,e^{-\m\r})L_{-1}+\frac{C_{3}}{3}L_{0}+\frac{i}{6}e^{\r\m}(C_{7}-C_{2})L_{1}\nn\\
&&-\frac{1}{2}e^{-2\r\m}(\tilde\cj_{UV}^{4}-iC_{5})W_{-2}+e^{-\m\r}C_{1}W_{-1}-(\sqrt{3} C_{8}-\frac{3}{4}C_{3})W_{0}\nn\\
&&-
\frac{i}{4}e^{\m\r}(C_{2}+C_{7})W_{1}-\frac{1}{8}e^{2\r\m}(\tilde\cj_{UV}^{4}+iC_{5})W_{2}+\frac{i}{2\sqrt{3}}\tilde\cj_{UV}^{9}\, J\bigg)\,dx^+\,.\label{2prundown}
\end{eqnarray}
This result should be matched with the bosonic part of the perturbed asymptotic AdS connection \eqref{ggcnt0}, which we rewrite here
\begin{equation}\label{a13tbcmp}
\G=\big(L_1e^{\r}+\cl_{IR} e^{-\r} L_{-1}+\cj_{IR} J+\ca_{IR} e^{-\r}K_{-1}+\cw_{IR} e^{-2\r}W_{-2}\big) dx^{+}+L_{0}d\r\,.
\end{equation}
By comparing \eqref{2prundown} and \eqref{a13tbcmp}, we see that the non-highest-weight components, which are the coefficients of the $K_{1},~K_{0},~L_{0},~W_{2,1,0,-1}$ generator, vanish. This fixes all the integration constants in \eqref{2prundown} as
\begin{equation}
C_{1}=C_{2}=C_{3}=C_{7}=C_{8}=0,~C_{5}=i\tilde\cj_{UV}^{4}\,.
\end{equation}
Under this parametrization, the perturbed interpolation solution \eqref{2pintuv} reduces to
\begin{eqnarray}
\G^{(2p)}&=& \bigg( \frac{1}{3}(4 e^{\m\r}- e^{e^{\r}+\m\r}) L_{1}- (e^{\m\r}-e^{e^{\r}+\m\r}) K_{1}\bigg)dx^{+}+(\m L_{0}+e^{\r}\hhl_{0})d\r \nn\\
&&+
\bigg((-e^{-e^\r-\m\r}\tilde{\cl}_{UV}-\tilde\cj_{UV}^{6} e^{-\m\r})\hhj_1+i(-e^{-e^\r-\m \r}\tilde{\cl}_{UV}-\tilde\cj_{UV}^{6} e^{-\m\r})\hhj_2\nn\\
&&+
\tilde\cj_{UV}^{4} e^{-2\m\r}\hhj_4+\tilde{\cl}_{UV}e^{-e^\r+\m\r} \hhl_{-1}
+(e^{-e^\r-\m\r}\tilde{\cl}_{UV}+\tilde\cj_{UV}^{6} e^{-\m\r})\hhj_6\nn\\
&&+i(e^{-e^\r-\m\r}\tilde{\cl}_{UV}+\tilde\cj_{UV}^{6} e^{-\m\r})\hhj_7
+i\,\tilde\cj_{UV}^{4} e^{-2\m\r}\hhj_5+\tilde\cj_{UV}^{9}\, \hhj_9\bigg)\,dx^+ \,.\label{2pintuvset}
\end{eqnarray}
The UV limit of this perturbed interpolation is achieved at $\r \to \infty$, $\m=0$
\begin{eqnarray}
\G^{(2p)}_{UV}&=& \hhl_1 e^{e^{\r}}dx^++e^{\r}\hhl_{0}d\r +
\big(-\tilde\cj_{UV}^{6} \hhj_1-i\,\tilde\cj_{UV}^{6} \hhj_2+\tilde\cj_{UV}^{4} \hhj_4+i\,\tilde\cj_{UV}^{4} \hhj_5\nn\\
&&+
\tilde\cj_{UV}^{6} \hhj_6+i\tilde\cj_{UV}^{6} \hhj_7
+\tilde\cj_{UV}^{9}\, \hhj_9\big)dx^+\,.
\end{eqnarray}
And the IR limit of this perturbed interpolation \eqref{2pintuvset}, which is ran down from the UV connection, reduces to
\begin{eqnarray}
\G^{(2p)}_{IR}&=& e^{\m\r} L_{1}dx^{+}+\m L_{0}d\r +
\bigg(e^{-\m \r}(\tilde{\cl}_{UV}+\tilde\cj_{UV}^{6})K_{-1}-\frac{1}{3}e^{-\m \r}(\tilde{\cl}_{UV}+4\tilde\cj_{UV}^{6})L_{-1}\nn\\
&&-e^{-2\r\m}\tilde\cj_{UV}^{4}W_{-2}+\frac{i}{2\sqrt{3}}\tilde\cj_{UV}^{9}\, J\bigg)dx^+\,. \end{eqnarray}
A further field redefinition $\r\to\tilde{\r}=\m\r$ put it to the conventional form
\begin{eqnarray}
\G^{(2p)}_{IR}&=& e^{\tilde \r} L_{1}dx^{+}+ L_{0}d\tilde \r +
\bigg(e^{-\tilde \r}(\tilde{\cl}_{UV}+\tilde\cj_{UV}^{6})K_{-1}-\frac{1}{3}e^{-\tilde \r}(\tilde{\cl}_{UV}+4\tilde\cj_{UV}^{6})L_{-1}\nn\\
&&-e^{-2\tilde \r}\tilde\cj_{UV}^{4}W_{-2}+\frac{i}{2\sqrt{3}}\tilde\cj_{UV}^{9}\, J\bigg)dx^+\,.\label{2prundowncmp}
\end{eqnarray}
Comparing to \eqref{a13tbcmp}, with the trivial identification $\tilde{\r}=\r$, we get the relations
\begin{subequations}\label{uvdownir}
\begin{eqnarray}
  \cl_{IR}&=&-\frac{1}{3}\tilde{\cl}_{UV}-\frac{4}{3}\tilde\cj_{UV}^{6} \\
  \ca_{IR}&=&\tilde{\cl}_{UV}+\tilde\cj_{UV}^{6} \\
  \cw_{IR}&=& -\tilde{\cj}_{UV}^4\\
  \cj_{IR}&=&\frac{i}{2\sqrt{3}}\tilde{\cj}_{UV}^9\,.
  \end{eqnarray}
\end{subequations}
Thus we have established relations between the UV operators and the IR operators and the relations do not depend on the parameter $\m$. The spacetime spin-3 IR current $\cw_{IR}$ is locally related to the spacetime spin-1 UV current, and the spacetime spin-2 IR currents $\cl_{IR},\, \ca_{IR}$ are related to certain combination of the spacetime  spin-$2, 1$ currents $\tilde{\cl}_{UV},~\tilde\cj_{UV}^{6}$. The R-current $\cj_{IR}$ in the IR theory is related to $\tilde{\cj}_{UV}^9$. From  the CFT point of view, the conformal dimensions of the UV operators $\tilde{\cl}_{UV},\tilde{\cj}_{UV}^4,\tilde{\cj}_{UV}^6,\tilde{\cj}_{UV}^9$ becomes $2, \, 3,\, 2,\, 1$ in the IR. The most significant difference between the current case and the previous case (also the bosonic case \cite{agkp}) is that the spin-$2$ $\tilde{\cl}_{UV}$ field does not acquire anomalous dimension. Instead, the dimension of the spin-1 current $\tilde{\cj}_{UV}^4$ jumps by 2. We wish to give a detailed analysis of this interesting  phenomena in the future.

$\bullet$ {\bf An alternative approach to get the relation \eqref{uvdownir}}\\
The above solution is obtained by running from the perturbed UV vacuum to the IR theory around the interpolation. Alternatively, we can start from the perturbed IR vacuum and go to match with the perturbed UV vacuum. Since the interpolation solution is a pure algebraic expression, the result of the matching should not depend on the direction we go along the interpolation. We will illustrate this now.

We start by perturbing the interpolation \eqref{2p} with IR fields \eqref{a13tbcmp}
\begin{eqnarray}
\G^{(2p)}&=& \bigg( \frac{1}{3}(4 e^{\m\r}- e^{e^{\r}+\m\r}) L_{1}- (e^{\m\r}-e^{e^{\r}+\m\r}) K_{1}\bigg)dx^{+}+(\m L_{0}+e^{\r}\hhl_{0})d\r \nn\\
&&+  (\cl'_{IR}L_{-1}+\cj'_{IR} J+\ca'_{IR} K_{-1}+\cw'_{IR} W_{-2})dx^+\,,
\end{eqnarray}
where again all fields depend on $(x^+,x^-,\r)$. Solving the equation of motion gives
\begin{eqnarray}
\G^{(2p)}&=& \big( \frac{1}{3}(4 e^{\m\r}- e^{e^{\r}+\m\r}) L_{1}- (e^{\m\r}-e^{e^{\r}+\m\r}) K_{1}\big)dx^{+}+(\m L_{0}+e^{\r}\hhl_{0})d\r \nn\\
&&+  \bigg(\frac{1}{9} e^{-\r \m  } \big(e^{-e^\r} (-4 \ca_{IR}-3 \cl_{IR})+4 (\ca_{IR}+3 \cl_{IR})\big)L_{-1}+\cw_{IR}e^{-2 \r \m} W_{-2}\nn\\
&&+\cj_{IR} J-\frac{1}{3} e^{-e^\r-\r \m } \big(-4 \ca_{IR}-3 \cl_{IR}+e^{e^\r} (\ca_{IR}+3 \cl_{IR})\big) K_{-1}\bigg)dx^+\,.
\end{eqnarray}
We run this interpolation to the UV theory by taking $\r\to \infty,~\m=0$ followed by a redefinition $\tilde\r=e^{\r}$. We then rewrite this result in the set of bases of the UV theory
\begin{eqnarray}
\G^{(2p)}_{UV}&=& e^{\tilde \r}\hhl_{1}dx^{+}+\hhl_{0}d\tilde\r +\bigg((\frac{\ca_{IR}}{3}+ \cl_{IR})\hhj_{1}+i (\frac{\ca_{IR}}{3}+ \cl_{IR})\hhj_{2}\nn\\
\nn&&-\cw_{IR}\hhj_{4}-i\cw_{IR}\hhj_{5}-(\frac{\ca_{IR}}{3}+ \cl_{IR})\hhj_{6}-i (\frac{\ca_{IR}}{3}+ \cl_{IR})\hhj_{7}-2\sqrt{3}i \cj_{IR}\hhj_{9} \\
&&+(\frac{4\ca_{IR}}{3}+ \cl_{IR})e^{-\tilde \r}\hhl_{-1}\bigg)dx^+\,.
\end{eqnarray}
Comparing this with the perturbed UV connection in the ``highest weight'' gauge \eqref{uvpert}, we get the following matching result
\begin{subequations}
\begin{eqnarray}
\tilde{\cl}_{UV} &=&\frac{4}{3}\ca_{IR}+\cl_{IR}\,, \\
\tilde{\cj}_{UV}^4  &=&-\cw_{IR}\,,\\
\tilde\cj_{UV}^{6} &=& -(\frac{1}{3}\ca_{IR}+\cl_{IR})\,,\\
\tilde{\cj}_{UV}^9 &=&-{2\sqrt{3}}i{\cj}_{IR}\,.
\end{eqnarray}
\end{subequations}
This agrees with \eqref{uvdownir}, which shows the  consistency of our computations.

Another observation from our computations in both the two cases, namely (i) matching the fields in  AdS$^{(1)}$ with fields in AdS$^{(\rm p)}$ and (ii) matching the fields in  AdS$^{(2)}$ with fields in AdS$^{(\rm p)}$, is that not any perturbation around the AdS vacuum is compatible with the RG flow. This is reflected in the fact that only a subset of the UV fields can be matched with IR fields in our analysis. The natural question to ask is why this is the case? Is this a limitation of our method or there are physical reasons behind? We hope to get back to this question in future work.

\subsubsection{Dual operators in the two UV theories}
As shown in the previous section, we relate the operators in the IR theory with the operators in the UV theories separately.  Then the natural question to ask is are there operators in each of the two UV theories that flow to the same IR operator? If so, we thus find operators in the two different theories dual to each other in the sense that they flow to the same IR operator.  This question is easy to answer given the matching result \eqref{1poptmap} and \eqref{uvdownir}, we find four pairs of operators that flow to the same IR operators. The result is summarized in Table \ref{optdual}.
\renewcommand{\arraystretch}{1.5}
\begin{table}[htdp]
\begin{center}
\begin{tabular}{|ccc|c|}
\hline
AdS$^{(1)}$ operators (UV)& $\bigg|$& AdS$^{(2)}$ operators (UV)& AdS$^{(\rm p)}$ operators (IR)\\\hline\hline
$\hj^{1}_{UV}+\hj^{4}_{UV}+3\hj^{5}_{UV},~\big\{\,\hat\cl_{UV}\,\big\}$ &$\sim$& $\tilde{\cj}_{UV}^{6}$& $\ca_{IR}+3\cl_{IR}$\\\hline
$\cm_{UV}-\cn_{UV}$&$\sim$&$\tilde{\cj}_{UV}^{4}$& $\cw_{IR}$\\\hline
\big\{\,$\hat\cj_{UV}^{2}\,\big\}$&$\sim$&$4\tilde{\cj}_{UV}^{6}+\tilde\cl_{UV}$& $\cl_{IR}$\\\hline
\big\{\,$\hat\cj_{UV}^{1}+\hat\cj_{UV}^{4}\,\big\}$&$\sim$&$\tilde{\cj}_{UV}^{9}$& $\cj_{IR}$\\\hline
\end{tabular}
\end{center}
\caption{This table lists the pairs of UV operators in the two different UV theories. When the UV operators are  related to the IR operators through identities with derivatives involved, we put the UV operators in the curly bracket $\big\{\,\cdots\,\big\}$. }
\label{optdual}
\end{table}%
Note that we do not direct set the UV operators equal to each other, since they are not directly related to each other. Rather, they are related by the same IR operators they flow into. In addition,  some UV operators are related to the IR operators through identities with derivatives,\fn{This may be considered as relations between descendent fields.} these type of UV operators are put in the curly bracket $\big\{\,\cdots\,\big\}$.  Another observation that is interesting is that for some pairs of the dual operators, their conformal dimension are different. This is not surprising since when run down to the IR theory, many operators acquire anomalous dimensions, as shown in  section \ref{optsec1p} and \ref{optsec2p}. This dual relation can be further studied in the content of the dual CFT.

\subsection{A remark on the central charge}
The general form of the central charge in the bosonic case is shown \cite{cfpt} to be
\begin{equation}\label{ccgen}
  c=12 \, k_{CS}\, \mbox{Tr}(L_0^2)\,.
\end{equation}
From our computation, we see the general form of the central charge in the supersymmetric case is
\begin{equation}\label{ccgens}
  c=12 \, k_{CS}\, \big|\mbox{STr}(L_0^2)\big|\,,
\end{equation}
where $k_{CS}$ is the Chern-Simons level. From our explicit computation (or \eqref{ccgens}), we see that the central charge of the IR theory, corresponding to the principal embedding, is larger than the central charges of the UV theories. This is similar to what happened in the bosonic case \cite{agkp}, and is not a violation of the $c$-theorem since the RG flow is triggered by Lorentz violating operator deformations.

Further notice that in both the two non-principal embeddings AdS$^{(1)}$ and AdS$^{(2)}$, there are always some spin-$1$ generators whose commutators have negative central terms \eqref{jj1}, \eqref{xx1}, \eqref{jj2}. This observation implies that both the two non-principal embeddings are dual to non-unitary CFTs, according to the analysis in \cite{chl}. Besides, the argument to circumvent this non-unitarity in \cite{aggrr2} may be extended to the supersymmetric case as well.

\section{Discussion}

\subsection{Relation with the Hamiltonian reductions of the WZW model}

A relevant question is how to understand these different embeddings holographically.  It has been shown, e.g. in \cite{drs}, that the resulting theories obtained by classical Hamiltonian reduction of the WZW model based on some supergroup possesses super-W symmetry. In addition,  it is shown in \cite{ais} that there are three different Hamiltonian reductions of the  current algebra associated with the Lie superalgebra $sl(3|2)$, each containing the usual Virasoro algebra as a subalgebra.\fn{There are actually 5 different reductions, but two of them has either constrained stress supercurrent or a spin-0 current, both of which seem unusual and are not considered here \cite{ais}.}

One of the reductions in \cite{ais} gives the $\cn=2$ super-W$_{3}$ algebra. As discussed in section \ref{emb1}, the  asymptotic symmetry of our principal embedding is precisely the  super-W$_{3}$ algebra. The second reduction in \cite{ais} gives rise to a $u(2|1)$ nonlinear superconformal algebra, which is the same algebra found in \cite{dth} . As discussed in section \ref{nonem1}, this algebra matches with the asymptotic symmetry algebra of the non-principal embedding I. The third reduction in \cite{ais} gives rise to a $u(3)$ nonlinear superconformal algebra, which is the same algebra found in \cite{bershadsky}. As discussed in section \ref{emb2}, this algebra matches with the asymptotic symmetry algebra of the non-principal embedding II.
Thus we see an exact match between
\begin{enumerate}
\item the asymptotic symmetry algebras of the higher-spin theories corresponding to  the three different embeddings of $osp(1|2)$ into the $sl(3|2)$ superalgebra and
\item the resulting algebra from the three different Hamiltonian reductions of the Lie superalgebra $sl(3|2)$.
\end{enumerate}
The work \cite{ais} was done purely algebraically and does not depend on the field theory realization. So it is reasonable to believe that the structure of the three different reductions in \cite{ais} should also be present in the Hamiltonian reductions of the WZW model based on the supergroup $sl(3|2)$.

The close relation between the classical Hamiltonian reduction of some Lie (super)algebra and the asymptotic symmetry algebra of the higher-spin theory based on the same Lie (super)algebra has been known in the literature, see e.g. \cite{cfpt, gh,gghr, hp}. However, our  result contains more information in the sense that we give a physical interpretation to the different Hamiltonian reductions.  The above matching relates a certain Hamiltonian reduction to the IR theory and two other Hamiltonian reductions to the UV theories. Since each Hamiltonian reduction is achieved by imposing a set of constraints on the $sl(3|2)$ current algebra, we further relate different sets of constraints to the WZW model with different higher-spin theories in the UV and in the IR, respectively. It turns out that the constraints corresponding to each UV theory form a subset of the constraints corresponding to the IR theories. This correspondence can be shown in the Figure \ref{topdown} and Figure \ref{bottom}.
\begin{figure}[t]
\begin{picture}(320,35)(0,5)
\setlength{\unitlength}{.5mm}
\put(60,60){$sl(3|2)$}
\put(32,86){\small\bf Asymptotic symmetries}
\put(10,78){\scriptsize (from different $osp(1|2)$ embeddings into $sl(3|2)$)}
\put(60,52){\vector(-1, -1){15}}
\put(80,52){\vector(1, -1){15}}
\put(70,52){\vector(0,-1){38}}
\put(25,27){$u(2|1)$ SCA}
\put(25,20){\footnotesize(via $sl(2|1)$)}
\put(90,27){$u(3)$ SCA}
\put(90,20){\footnotesize(via $sl(1|2)$)}
\put(58,6){$\cn=2~ \cw_3$}
\put(58,-2){\footnotesize(via $sl(3|2)$)}
\multiput(156,0)(0,4){22}{\line(0,1){2}}
\put(196,86){\small\bf Hamiltonian reduction}
\put(170,78){\scriptsize (from different constraints on the $sl(3|2)$ currents)}
\put(218,60){$sl(3|2)$}
\put(220,52){\vector(-1, -1){15}}
\put(240,52){\vector(1, -1){15}}
\put(230,52){\vector(0,-1){36}}
\put(185,25){$u(2|1)$-SCA}
\put(250,25){$u(3)$-SCA}
\put(218,5){$\cn=2~ \cw_{3}$}
\end{picture}\\
[1mm]
  \caption{This figure shows the similarity between structures of the three different $osp(1|2)$ embeddings into the $sl(3|2)$ (left) and the three different Hamiltonian reductions of the $sl(3|2)$ superalgebra (right). The ``SCA" in both figures stands for ``superconformal algebra".}\label{topdown}
\vspace{-15mm}
\begin{picture}(320,60)(0,5)
\setlength{\unitlength}{.5mm}
\put(32,86){\small\bf Asymptotic symmetries}
\put(46,78){\scriptsize (related by RG flows)}
\put(65.5,15){\vector(1, -2){.5}}
\multiput(48,50)(1,-2){17}{\circle*{.1}}
\put(88.5,15){\vector(-1, -2){.5}}
\multiput(106,50)(-1,-2){17}{\circle*{.1}}
\put(25,63){$u(2|1)$ SCA}
\put(23,55){\footnotesize (UV theory I)}
\put(90,63){$u(3)$ SCA}
\put(90,55){\footnotesize
(UV theory II)}
\put(60,5){$\cn=2~ \cw_3$}
\put(60,-2){\footnotesize
(IR theory)}
\multiput(156,0)(0,4){22}{\line(0,1){2}}
\put(196,86){\small\bf Hamiltonian reduction}
\put(176,78){\scriptsize (related by secondary Hamiltonian reductions)}
\put(200,50){\vector(1, -2){15}}
\put(270,50){\vector(-1, -2){15}}
\put(185,55){$u(2|1)$-SCA}
\put(250,55){$u(3)$-SCA}
\put(220,5){$\cn=2~ \cw_{3}$}
\end{picture}\\
[1mm]
  \caption{The figure on the left shows the physical interpretation of the relations between the three embeddings. The dotted arrows represent RG flows. The figure on the right shows that the $\cn=2$ $\cw_3$ algebra can be obtained from secondary Hamiltonian reductions of the $u(3)$-SCA and $u(2|1)$-SCA. The three objects in each diagram are the same as those shown up in Figure \ref{topdown}. }\label{bottom}
\end{figure}

We can further establish a close relation between turning on some currents to trigger the RG flow in the higher-spin theory with putting more constraints to induce a further (secondary) Hamiltonian reduction. It is interesting to understand this correspondence in the future. Concretely, it is interesting to see how do the extra constraints imposed by the secondary Hamiltonian reduction \cite{ais} translate into the operators triggering the RG flow? If this translation is understood, can we apply it to the initial constraints put on the $sl(3|2)$?

Another motivation of this comparison between our results and the Hamiltonian reductions is the attempt to find Lagrangian descriptions of the CFT dual to the spin-3 supergravity with different embeddings. For the principal embedding, the dual CFT is proposed to be the $\bc \text{P}^{n}$ minimal models \cite{chr, cg,cg2} from the Kazama-Suzuki coset construction \cite{ks}. For the non-principal embeddings, we identify the symmetry algebras as two different non-linear extended superconformal algebras. But we do not have concrete realizations of these algebras at hand. As known for the bosonic  case,\fn{We thank Thomas Hartman for pointing this out to the author.} even with a special choice of the parameter in a Toda field theory, which can  be obtained from the Hamiltonian reduction of a WZW model, such that the Virasoro central charge of the  Toda theory agrees with the central charge of a certain $W_{n}$-minimal model, the Toda theory is not the Lagrangian description of the $W_{n}$-minimal model \cite{mansfield, ms, bs}. The obstruction is the mismatch of the spectrum: the minimal model contains more states than the Toda theory and certain projections are required for the matching. To overcome this difficulty, Mansfield and Spence suggested to consider the conformally extended Toda field theories instead of the original ones \cite{mansfield, ms}. We anticipate that the situation could be similar in our supersymmetric picture and a similar extension may be needed as well. It is interesting to work this out in the future and this may develop our understanding of the higher-spin holography.

\subsection{$osp(1|2)$ embedding and $sl(1|2)$ embedding}
The $osp(1|2)$ superalgebra can be regarded as the $\cn=1$ supersymmetric extension of the bosonic $sl(2)$ algebra, while $sl(2|1)$ is the $\cn=2$ supersymmetric extension. Our discussion is in the context of $\cn=2$ supersymmetric higher-spin theory. Then a question arises: why do we consider the $osp({1|2})$ embedding instead of the $sl(1|2)$ embedding?

It is shown in \cite{frs,rss} that any $sl(1|2)$ embedding provides an $osp(1|2)$ embedding. Furthermore, for the Lie superalgebra $sl(n|n-1)$, the $osp(1|2)$ embedding classifies the $sl(1|2)$ embedding.  Therefore considering the $osp(1|2)$ embedding in our case is the same as considering $sl(1|2)$ embedding.  We use the $osp(1|2)$ embedding since it is the simplest supersymmetric extension of the $sl(2)$ algebra and we do not need to further group the $\cn=1$ multiplets into $\cn=2$ multiplets.

\subsection{Generalizations to arbitrary superalgebra $sl(n+1|n)$}
In this paper, we have given a detailed analysis of the 3 different embeddings of $osp(1|2)$ into $sl(3|2)$.  One of them can be interpreted as an IR theory and the other two as UV theories. We find the two UV theories flow to the same IR theory. We can  generalize this analysis to higher rank algebras, say $sl(4|3)$,  and study the relations between the different embeddings. It is very possible that the principal embedding will again give a theory at IR,\fn{This is because the principal embedding leads to minimal number of primary fields after imposing the boundary conditions and gauge fixing conditions.} then the question left is that do all the different non-principal embeddings flow to this same theory? Are there ``cascade" scenarios, namely, are there embeddings corresponding to intermediate scales so that we can construct successive RG flows from some embedding A to some other embedding B then to the IR theory? In the present $sl(3|2)$ case, the structure is that the two non-principal embeddings give two UV theories flowing to the IR respectively. Our analysis in section \ref{uvuv} rules out the possibility of one UV theory being at an intermediate scale and of constructing two successive RG flows. However, the latter case is possible for higher-rank algebras. It is interesting to see what are the structures for general (bosonic or supersymmetric) higher-spin theories.
No matter what is the answer, there could be some interesting relations/dualities appear and this construction may provide a playground for further studies.

\section*{ Acknowledgement}

I would like to thank especially Thomas Hartman for multiple enlightening communications. I am grateful to Matthias Gaberdiel for his comments on the draft of this paper and illuminating suggestions. I appreciate the helpful discussions with Henriette Elvang, Yu-tin Huang and Mukund Rangamani.  CP is supported by NSF Grant PHY-0953232 and in part by the DoE Grant DE-SC0007859
 .

\appendix

\section{Structure constants of  $u(2|1)$ algebra }\label{strcnt}
The structure constants of the $u(2|1)$ algebra that show up in the non-principal embedding I in section \ref{non1} are listed here
\begin{eqnarray*}
  \nn &&f^{122} =f^{313}=f^{324}=f^{231}=f^{433}=f^{242}= 1\,,\\
  && f^{133}=f^{212}=f^{321}=f^{422}=f^{234}=f^{343}= -1 \,,\\
  && g^{111}=g^{133}=g^{252}=g^{231}=g^{433}=g^{544}=1\,,\\
  && g^{151}=g^{212}=g^{224}=g^{533}=g^{342}=g^{444}=-1\,,\\
  \nn && \g^{211}=\g^{215}=\g^{412}=\g^{142}=\g^{121}=\g^{125}=\g^{323}=\g^{233}=\g^{434}=\g^{435}=\g^{344}=\g^{345}=1\,,\\
  &&\q^{111}=\q^{312}=\q^{221}=\q^{422}=1\,,\\
    &&\bar\q^{111}=\bar\q^{212}=\bar\q^{321}=\bar\q^{422}=-1\,,
\end{eqnarray*}
And the coefficients of the central terms are
\begin{eqnarray*}
  \nn &&h^{11} = h^{44}=2\,,\\
  && h^{14}=h^{41}=h^{23}=h^{32}=1\,,\\
  && h^{15}=h^{51}=h^{45}=h^{54}=-1\,,\\
  && \b^{12}= \b^{34}=1\,\\
  && \b^{21}= \b^{43}=-1\,.
\end{eqnarray*}


\begin{thebibliography}{99}
\bibitem{Fronsdal}
  C.~Fronsdal,
  ``Massless Fields with Integer Spin,''  Phys.\ Rev.\ D {\bf 18}, 3624 (1978).

\bibitem{vasiliev}
  E.~S.~Fradkin and M.~A.~Vasiliev,
  ``On the Gravitational Interaction of Massless Higher Spin Fields,''
  Phys.\ Lett.\ B {\bf 189}, 89 (1987).
  M.~A.~Vasiliev,
  ``Higher Spin Algebras And Quantization On The Sphere And Hyperboloid,''
  Int.\ J.\ Mod.\ Phys.\ A {\bf 6}, 1115 (1991).
  M.~A.~Vasiliev,
  ``Higher spin gauge theories in four-dimensions, three-dimensions, and two-dimensions,''
  Int.\ J.\ Mod.\ Phys.\ D {\bf 5}, 763 (1996)
  [hep-th/9611024].
  M.~A.~Vasiliev,
  ``Higher spin gauge theories: Star product and AdS space,''
  In *Shifman, M.A. (ed.): The many faces of the superworld* 533-610
  [hep-th/9910096].
  

\bibitem{klp}
  I.~R.~Klebanov and A.~M.~Polyakov,
  ``AdS dual of the critical O(N) vector model,''  Phys.\ Lett.\ B {\bf 550}, 213 (2002)  [hep-th/0210114].  

\bibitem{ss}
  E.~Sezgin and P.~Sundell,
  ``Holography in 4D (super) higher spin theories and a test via cubic scalar couplings,''  JHEP {\bf 0507}, 044 (2005)  [hep-th/0305040].  

\bibitem{sundborg}
  B.~Sundborg,
  ``Stringy gravity, interacting tensionless strings and massless higher spins,''  Nucl.\ Phys.\ Proc.\ Suppl.\  {\bf 102}, 113 (2001)  [hep-th/0103247].  


  \bibitem{ss2}
  E.~Sezgin and P.~Sundell,
  ``Massless higher spins and holography,''
  Nucl.\ Phys.\ B {\bf 644}, 303 (2002)
  [Erratum-ibid.\ B {\bf 660}, 403 (2003)]
  [hep-th/0205131].
 


\bibitem{gmptwy}
  S.~Giombi, S.~Minwalla, S.~Prakash, S.~P.~Trivedi, S.~R.~Wadia and X.~Yin,
  ``Chern-Simons Theory with Vector Fermion Matter,''  Eur.\ Phys.\ J.\ C {\bf 72}, 2112 (2012)  [arXiv:1110.4386 [hep-th]].  

\bibitem{cmsy}
  C.~-M.~Chang, S.~Minwalla, T.~Sharma and X.~Yin,
  ``ABJ Triality: from Higher Spin Fields to Strings,''  arXiv:1207.4485 [hep-th].  
  
\bibitem{agy}
  O.~Aharony, G.~Gur-Ari and R.~Yacoby,
  ``d=3 Bosonic Vector Models Coupled to Chern-Simons Gauge Theories,''  JHEP {\bf 1203}, 037 (2012)  [arXiv:1110.4382 [hep-th]].  

\bibitem{mz1}
  J.~Maldacena and A.~Zhiboedov,
  ``Constraining Conformal Field Theories with A Higher Spin Symmetry,''  arXiv:1112.1016 [hep-th].  

\bibitem{mz2}
  J.~Maldacena and A.~Zhiboedov,
  ``Constraining conformal field theories with a slightly broken higher spin symmetry,''  arXiv:1204.3882 [hep-th].  

\bibitem{agy2}
  O.~Aharony, G.~Gur-Ari and R.~Yacoby,
  ``Correlation Functions of Large N Chern-Simons-Matter Theories and Bosonization in Three Dimensions,''  arXiv:1207.4593 [hep-th].  

\bibitem{aggmy}
  O.~Aharony, S.~Giombi, G.~Gur-Ari, J.~Maldacena and R.~Yacoby,
  ``The Thermal Free Energy in Large N Chern-Simons-Matter Theories,''  arXiv:1211.4843 [hep-th].  


\bibitem{dj}
  S.~R.~Das and A.~Jevicki,
  ``Large N collective fields and holography,''
  Phys.\ Rev.\ D {\bf 68}, 044011 (2003)
  [hep-th/0304093].
  

\bibitem{jjy}
  A.~Jevicki, K.~Jin and Q.~Ye,
  ``Collective Dipole Model of AdS/CFT and Higher Spin Gravity,''
  J.\ Phys.\ A {\bf 44}, 465402 (2011)
  [arXiv:1106.3983 [hep-th]].
  

\bibitem{dmr}
  M.~R.~Douglas, L.~Mazzucato and S.~S.~Razamat,
  ``Holographic dual of free field theory,''
  Phys.\ Rev.\ D {\bf 83}, 071701 (2011)
  [arXiv:1011.4926 [hep-th]].
 

  \bibitem{v12}
  M.~A.~Vasiliev,
  ``Holography, Unfolding and Higher-Spin Theory,''
  arXiv:1203.5554 [hep-th].
  


\bibitem{ahs}
  D.~Anninos, T.~Hartman and A.~Strominger,
  ``Higher Spin Realization of the dS/CFT Correspondence,''  arXiv:1108.5735 [hep-th].  

\bibitem{ns}
  G.~S.~Ng and A.~Strominger,
  ``State/Operator Correspondence in Higher-Spin dS/CFT,''  arXiv:1204.1057 [hep-th].  

\bibitem{ddjy}
  D.~Das, S.~R.~Das, A.~Jevicki and Q.~Ye,
  ``Bi-local Construction of Sp(2N)/dS Higher Spin Correspondence,''  arXiv:1205.5776 [hep-th].  

\bibitem{gy}
  S.~Giombi and X.~Yin,
  ``The Higher Spin/Vector Model Duality,''  arXiv:1208.4036 [hep-th].  


\bibitem{gg}
  M.~R.~Gaberdiel and R.~Gopakumar,
  ``An AdS$_3$ Dual for Minimal Model CFTs,''  Phys.\ Rev.\ D {\bf 83}, 066007 (2011)  [arXiv:1011.2986 [hep-th]].  

\bibitem{gg12}
  M.~R.~Gaberdiel and R.~Gopakumar,
  ``Triality in Minimal Model Holography,''
  JHEP {\bf 1207}, 127 (2012)
  [arXiv:1205.2472 [hep-th]].
  

\bibitem{ppr}
  E.~Perlmutter, T.~Prochazka and J.~Raeymaekers,
  ``The semiclassical limit of W$_N$ CFTs and Vasiliev theory,''
  arXiv:1210.8452 [hep-th].
  

\bibitem{cgkv}
  C.~Candu, M.~R.~Gaberdiel, M.~Kelm and C.~Vollenweider,
  ``Even spin minimal model holography,''
  arXiv:1211.3113 [hep-th].
  

\bibitem{chr}
  T.~Creutzig, Y.~Hikida and P.~B.~Ronne,
  ``Higher spin AdS$_3$ supergravity and its dual CFT,''
  arXiv:1111.2139 [hep-th].
  

\bibitem{chr1}
  T.~Creutzig, Y.~Hikida and P.~B.~Ronne,
  ``N=1 supersymmetric higher spin holography on AdS$_3$,''  arXiv:1209.5404 [hep-th].  

\bibitem{cg}
  C.~Candu and M.~R.~Gaberdiel,
  ``Supersymmetric holography on $AdS_3$,''
  arXiv:1203.1939 [hep-th].
  


\bibitem{cg2}
  C.~Candu and M.~R.~Gaberdiel,
  ``Duality in N=2 minimal model holography,''
  arXiv:1207.6646 [hep-th].
  

\bibitem{gghr}
  M.~R.~Gaberdiel, R.~Gopakumar, T.~Hartman and S.~Raju,
  ``Partition Functions of Holographic Minimal Models,''
  JHEP {\bf 1108}, 077 (2011)
  [arXiv:1106.1897 [hep-th]].
  

  \bibitem{hr}
  M.~Henneaux and S.~-J.~Rey,
  ``Nonlinear $W_{infinity}$ as Asymptotic Symmetry of Three-Dimensional Higher Spin Anti-de Sitter Gravity,''
  JHEP {\bf 1012}, 007 (2010)
  [arXiv:1008.4579 [hep-th]].
  

 \bibitem{cfpt}
  A.~Campoleoni, S.~Fredenhagen, S.~Pfenninger and S.~Theisen,
  ``Asymptotic symmetries of three-dimensional gravity coupled to higher-spin fields,''
  JHEP {\bf 1011}, 007 (2010)
  [arXiv:1008.4744 [hep-th]].
  %%CITATION = ARXIV:1008.4744;%

\bibitem{gh}
  M.~R.~Gaberdiel, T.~Hartman,
  ``Symmetries of Holographic Minimal Models,''
  JHEP {\bf 1105}, 031 (2011).
  [arXiv:1101.2910 [hep-th]].

\bibitem{hlpr}
  M.~Henneaux, G.~Lucena Gomez, J.~Park and S.~-J.~Rey,
  ``Super- W(infinity) Asymptotic Symmetry of Higher-Spin $AdS_3$ Supergravity,''
  JHEP {\bf 1206}, 037 (2012)
  [arXiv:1203.5152 [hep-th]].
  %%CITATION = ARXIV:1203.5152;%%


\bibitem{hp}
  K.~Hanaki and C.~Peng,
  ``Symmetries of Holographic Super-Minimal Models,''
  arXiv:1203.5768 [hep-th].
  %%CITATION = ARXIV:1203.5768;%%

\bibitem{ahn1}
  C.~Ahn,
  ``The Large N 't Hooft Limit of Kazama-Suzuki Model,''
  JHEP {\bf 1208}, 047 (2012)
  [arXiv:1206.0054 [hep-th]].
  %%CITATION = ARXIV:1206.0054;%%

\bibitem{ahn2}
  C.~Ahn,
  ``The Operator Product Expansion of the Lowest Higher Spin Current at Finite N,''
  arXiv:1208.0058 [hep-th].
  %%CITATION = ARXIV:1208.0058;%%

  \bibitem{chr2}
  T.~Creutzig, Y.~Hikida and P.~B.~Ronne,
  ``Three point functions in higher spin AdS$_3$ supergravity,''
  arXiv:1211.2237 [hep-th].
  %%CITATION = ARXIV:1211.2237;%%

  \bibitem{mz}
  H.~Moradi and K.~Zoubos,
  ``Three-Point Functions in N=2 Higher-Spin Holography,''
  arXiv:1211.2239 [hep-th].
  %%CITATION = ARXIV:1211.2239;%%

\bibitem{chjm}
  A.~Castro, E.~Hijano, A.~Lepage-Jutier and A.~Maloney,
  ``Black Holes and Singularity Resolution in Higher Spin Gravity,''
  JHEP {\bf 1201}, 031 (2012)
  [arXiv:1110.4117 [hep-th]].
  %%CITATION = ARXIV:1110.4117;%%

\bibitem{cggr}
  A.~Castro, R.~Gopakumar, M.~Gutperle and J.~Raeymaekers,
  ``Conical Defects in Higher Spin Theories,''
  JHEP {\bf 1202}, 096 (2012)
  [arXiv:1111.3381 [hep-th]].
  %%CITATION = ARXIV:1111.3381;%%

  \bibitem{gk}
  M.~Gutperle and P.~Kraus,
  ``Higher Spin Black Holes,''
  JHEP {\bf 1105}, 022 (2011)
  [arXiv:1103.4304 [hep-th]].
  %%CITATION = ARXIV:1103.4304;%%

\bibitem{kp}
  P.~Kraus and E.~Perlmutter,
  ``Partition functions of higher spin black holes and their CFT duals,''
  JHEP {\bf 1111}, 061 (2011)
  [arXiv:1108.2567 [hep-th]].
  %%CITATION = ARXIV:1108.2567;%%

  \bibitem{ghj}
  M.~R.~Gaberdiel, T.~Hartman and K.~Jin,
  ``Higher Spin Black Holes from CFT,''
  JHEP {\bf 1204}, 103 (2012)
  [arXiv:1203.0015 [hep-th]].
  %%CITATION = ARXIV:1203.0015;%%


  \bibitem{tan}
  H.~S.~Tan,
  ``Exploring Three-dimensional Higher-Spin Supergravity based on $sl(N |N - 1)$ Chern-Simons theories,''
  arXiv:1208.2277 [hep-th].
  %%CITATION = ARXIV:1208.2277;%%

\bibitem{dd}
  S.~Datta and J.~R.~David,
  ``Supersymmetry of classical solutions in Chern-Simons higher spin supergravity,''
  arXiv:1208.3921 [hep-th].
  %%CITATION = ARXIV:1208.3921;%%


\bibitem{ggr}
  M.~Gary, D.~Grumiller and R.~Rashkov,
  ``Towards non-AdS holography in 3-dimensional higher spin gravity,''
  JHEP {\bf 1203}, 022 (2012)
  [arXiv:1201.0013 [hep-th]].
  %%CITATION = ARXIV:1201.0013;%%

%\bibitem{Afshar:2012nk}
  H.~Afshar, M.~Gary, D.~Grumiller, R.~Rashkov and M.~Riegler,
  ``Non-AdS holography in 3-dimensional higher spin gravity - General recipe and example,''
  arXiv:1209.2860 [hep-th].
  %%CITATION = ARXIV:1209.2860;%%

%\bibitem{riegler}
  M.~Riegler,
  ``Asymptotic Symmetry Algebras in Non-Anti-de-Sitter Higher-Spin Gauge Theories,''
  arXiv:1210.6500 [hep-th].
  %%CITATION = ARXIV:1210.6500;%%

\bibitem{gg2}
  M.~R.~Gaberdiel and R.~Gopakumar,
  ``Minimal Model Holography,''
  arXiv:1207.6697 [hep-th].
  %%CITATION = ARXIV:1207.6697;%%

\bibitem{agkp1}
  M.~Ammon, M.~Gutperle, P.~Kraus and E.~Perlmutter,
  ``Black holes in three dimensional higher spin gravity: A review,''
  arXiv:1208.5182 [hep-th].
  %%CITATION = ARXIV:1208.5182;%%

    %\cite{agkp}
\bibitem{agkp}
  M.~Ammon, M.~Gutperle, P.~Kraus and E.~Perlmutter,
  ``Spacetime Geometry in Higher Spin Gravity,''
  JHEP {\bf 1110}, 053 (2011)
  [arXiv:1106.4788 [hep-th]].
  %%CITATION = ARXIV:1106.4788;%%

\bibitem{dfk}
  J.~R.~David, M.~Ferlaino and S.~P.~Kumar,
  ``Thermodynamics of higher spin black holes in 3D,''
  arXiv:1210.0284 [hep-th].


  \bibitem{mansfield}
  P.~Mansfield,
  ``Conformally extended Toda theories,''  Phys.\ Lett.\ B {\bf 242}, 387 (1990).  %%CITATION = PHLTA,B242,387;%%

\bibitem{ms}
  P.~Mansfield and B.~J.~Spence,
  ``Toda theories, the geometry of W algebras and minimal models,''  Nucl.\ Phys.\ B {\bf 362}, 294 (1991).  %%CITATION = NUPHA,B362,294;%%

 \bibitem{w88}
  E.~Witten,
  ``(2+1)-Dimensional Gravity as an Exactly Soluble System,''
  Nucl.\ Phys.\  {\bf B311}, 46 (1988).

%\cite{Witten:2007kt}
\bibitem{Witten:2007kt}
  E.~Witten,
  ``Three-Dimensional Gravity Revisited,''  arXiv:0706.3359 [hep-th].  %%CITATION = ARXIV:0706.3359;%%

\bibitem{fss}
  L.~Frappat, P.~Sorba and A.~Sciarrino,
  ``Dictionary on Lie superalgebras,''
  hep-th/9607161.
  %%CITATION = HEP-TH/9607161;%%

 \bibitem{fl}
  E.~S.~Fradkin and V.~Y.~.Linetsky,
  ``Supersymmetric Racah basis, family of infinite dimensional superalgebras, SU($\infty + 1|\infty$) and related 2-D models,''
  Mod.\ Phys.\ Lett.\ A {\bf 6}, 617 (1991).
  %%CITATION = MPLAE,A6,617;%%

  \bibitem{frs}
  L.~Frappat, E.~Ragoucy and P.~Sorba,
  ``W algebras and superalgebras from constrained WZW models: A Group theoretical classification,''
  Commun.\ Math.\ Phys.\  {\bf 157}, 499 (1993)
  [hep-th/9207102].
  %%CITATION = HEP-TH/9207102;%%

\bibitem{scheunert}
M.~Scheunert,
``Serre-type relations for special linear Lie superalgebras," Letters in Mathematical Physics, Volume 24, Issue 3, pp.173-181

\bibitem{dggt}
  F.~Delduc, F.~Gieres, S.~Gourmelen and S.~Theisen,
  ``Nonstandard matrix formats of Lie superalgebras,''  math-ph/9901017.  %%CITATION = MATH-PH/9901017;%%

%\cite{hms}
\bibitem{hms}
  M.~Henneaux, L.~Maoz and A.~Schwimmer,
  ``Asymptotic dynamics and asymptotic symmetries of three-dimensional extended AdS supergravity,''  Annals Phys.\  {\bf 282}, 31 (2000)  [hep-th/9910013].  %%CITATION = HEP-TH/9910013;%%

\bibitem{ito93}
  K.~Ito,
  ``Free field realization of N=2 superW(3) algebra,''  Phys.\ Lett.\ B {\bf 304}, 271 (1993)  [hep-th/9302039].  %%CITATION = HEP-TH/9302039;%%

  %\cite{dth}
\bibitem{dth}
  F.~Defever, W.~Troost and Z.~Hasiewicz,
  ``Superconformal algebras with quadratic nonlinearity,''  Phys.\ Lett.\ B {\bf 273}, 51 (1991).  %%CITATION = PHLTA,B273,51;%%

\bibitem{bershadsky}
  M.~A.~Bershadsky,
  ``Superconformal Algebras In Two-dimensions With Arbitrary N,''  Phys.\ Lett.\ B {\bf 174}, 285 (1986).  %%CITATION = PHLTA,B174,285;%%

\bibitem{mathieu}
  P.~Mathieu,
  ``Representation Of The So(n) And U(n) Superconformal Algebras Via Miura Transformations,''
  Phys.\ Lett.\ B {\bf 218}, 185 (1989).
  %%CITATION = PHLTA,B218,185;%%

\bibitem{lprsw}
  H.~Lu, C.~N.~Pope, L.~J.~Romans, X.~Shen and X.~J.~Wang,
  ``Polyakov construction of the N=2 superW(3) algebra,''  Phys.\ Lett.\ B {\bf 264}, 91 (1991).  %%CITATION = PHLTA,B264,91;%%

  \bibitem{ik}
  E.~Ivanov and S.~Krivonos,
  ``Superfield realizations of N=2 superW(3),''  Phys.\ Lett.\ B {\bf 291}, 63 (1992)  [Erratum-ibid.\ B {\bf 301}, 454 (1993)]  [hep-th/9204023].  %%CITATION = HEP-TH/9204023;%%



\bibitem{ahn93}
  C.~-h.~Ahn,
  ``Free superfield realization of N=2 quantum superW(3) algebra,''  Mod.\ Phys.\ Lett.\ A {\bf 9}, 271 (1994)  [hep-th/9304038].  %%CITATION = HEP-TH/9304038;%%

\bibitem{fgs}
  J.~-F.~Fortin, B.~Grinstein and A.~Stergiou,
  ``Scale without Conformal Invariance in Four Dimensions,''
  arXiv:1206.2921 [hep-th].
  %%CITATION = ARXIV:1206.2921;%%

%  \bibitem{fgs2}
  J.~-F.~Fortin, B.~Grinstein and A.~Stergiou,
  ``A generalized c-theorem and the consistency of scale without conformal invariance,''
  arXiv:1208.3674 [hep-th].
  %%CITATION = ARXIV:1208.3674;%%

\bibitem{lpr}
  M.~A.~Luty, J.~Polchinski and R.~Rattazzi,
  ``The $a$-theorem and the Asymptotics of 4D Quantum Field Theory,''
  arXiv:1204.5221 [hep-th].
  %%CITATION = ARXIV:1204.5221;%%

 \bibitem{chl}
  A.~Castro, E.~Hijano and A.~Lepage-Jutier,
  ``Unitarity Bounds in AdS$_3$ Higher Spin Gravity,''
  JHEP {\bf 1206}, 001 (2012)
  [arXiv:1202.4467 [hep-th]].
  %%CITATION = ARXIV:1202.4467;%%

\bibitem{aggrr2}
  H.~Afshar, M.~Gary, D.~Grumiller, R.~Rashkov and M.~Riegler,
  ``Semi-classical unitarity in 3-dimensional higher-spin gravity for non-principal embeddings,''
  arXiv:1211.4454 [hep-th].
  %%CITATION = ARXIV:1211.4454;%%

 \bibitem{drs}
  F.~Delduc, E.~Ragoucy and P.~Sorba,
  ``SuperToda theories and W algebras from superspace Wess-Zumino-Witten models,''  Commun.\ Math.\ Phys.\  {\bf 146}, 403 (1992).  %%CITATION = CMPHA,146,403;%%

\bibitem{ais}
  C.~-h.~Ahn, E.~Ivanov and A.~S.~Sorin,
  ``N=2 affine superalgebras and Hamiltonian reduction in N=2 superspace,''
  Commun.\ Math.\ Phys.\  {\bf 183}, 205 (1997)
  [hep-th/9508005].
  %%CITATION = HEP-TH/9508005;%%

 \bibitem{ks}
  Y.~Kazama and H.~Suzuki,
 ``New N=2 Superconformal Field Theories and Superstring Compactification,''
  Nucl.\ Phys.\ B {\bf 321}, 232 (1989).
  %%CITATION = NUPHA,B321,232;%%

\bibitem{bs}
  P.~Bouwknegt and K.~Schoutens,
  ``W symmetry in conformal field theory,''
  Phys.\ Rept.\  {\bf 223}, 183 (1993)
  [hep-th/9210010].
  %%CITATION = HEP-TH/9210010;%

 \bibitem{rss}
  E.~Ragoucy, A.~Sevrin and P.~Sorba,
 ``Strings from N=2 gauged Wess-Zumino-Witten models,''
  Commun.\ Math.\ Phys.\  {\bf 181}, 91 (1996)
  [hep-th/9511049].
  %%CITATION = HEP-TH/9511049;%%













\end{thebibliography}
\end{document}